\documentclass[aps,prd,notitlepage,nofootinbib,preprintnumbers,superscriptaddress,10pt]{revtex4-2}
\usepackage{here}
\usepackage{graphicx}
\usepackage{amsmath,amsthm,amssymb}
\usepackage{bm}
\usepackage{color}

\usepackage{amsfonts}
\usepackage{dcolumn}
\usepackage{hyperref}
\allowdisplaybreaks[1]
\usepackage{stackengine}

\newcommand{\be}{\begin{eqnarray}}
\newcommand{\ee}{\end{eqnarray}}

\newcommand{\bem}{\begin{bmatrix}}
\newcommand{\eem}{\end{bmatrix}}

\newcommand{\mn}{{\mu \nu}}

\newcommand{\mF}{\mathcal{F}}

\newcommand{\mh}{\mathfrak{h}}

\newcommand{\mm}[1]{\textcolor{cyan}{[MM:~#1]}}

\allowdisplaybreaks
\begin{document}

\title{Black hole thermodynamics in Horndeski theories}

\author{Masato Minamitsuji}
\affiliation{
Centro de Astrof\'{\i}sica e Gravita\c c\~ao - CENTRA, Departamento de F\'{\i}sica, Instituto Superior T\'ecnico - IST, Universidade de Lisboa - UL, Av.~Rovisco Pais 1, 1049-001 Lisboa, Portugal}

\author{Kei-ichi Maeda}
\affiliation{
Department of Physics, Waseda University, 3-4-1 Okubo, Shinjuku, Tokyo 169-8555, Japan}
\affiliation{Center for Gravitational Physics and Quantum Information, Yukawa Institute for Theoretical Physics, Kyoto University, 606-8502, Kyoto, Japan}

\begin{abstract}
We investigate thermodynamics of static and spherically symmetric black holes (BHs) in the Horndeski theories. Because of the presence of the higher-derivative interactions and the nonminimal derivative couplings of the scalar field, the standard Wald entropy formula may not be directly applicable. Hence, following the original formulation by Iyer and Wald, we obtain the differentials of the BH entropy and the total mass of the system in the Horndeski theories, which lead to the first-law of thermodynamics via the conservation of the Hamiltonian. Our formulation covers the case of the static and spherically symmetric BH solutions with the static scalar field and those with the linearly time-dependent scalar field in the shift-symmetric Horndeski theories. We then apply our results to explicit BH solutions in the Horndeski theories. In the case of the conventional scalar-tensor theories  and the Einstein-scalar-Gauss-Bonnet theories, we recover the BH entropy obtained by the Wald entropy formula. In the shift-symmetric theories, in the case of the BH solutions with the static scalar field we show that the BH entropy follows the ordinary area law even in the presence of the nontrivial profile of the scalar field. On the other hand, in the case of the BH solutions where the scalar field linearly depends on time, i.e., the stealth Schwarzschild and Schwarzschild-(anti-) de Sitter solutions, the BH entropy also depends on the profile of the scalar field. By use of the entropy, we find that there exists some range of the parameters in which Schwarzschild$-$(AdS) BH with non-trivial scalar field is thermodynamically stable than Schwarzschild$-$(AdS) BH without scalar field in general relativity. Finally,~we consider the Horndeski theories minimally coupled to the $U(1)$-invariant vector field where BH solutions contain the mass and the electric charge, and clarify the conditions under which the differential of the BH entropy is integrable in spite of the presence of the two independent charges.
\end{abstract}

\date{\today}


\maketitle

\section{Introduction}
\label{introsec}

General relativity (GR) is known as the unique gravitational theory in four dimensions which only contains two degrees of freedom (DOFs) of metric and preserves the Lorentz symmetry~\cite{Lovelock:1972vz}.
GR has been tested by the local experiments as well as the astrophysical probes \cite{Will:2014kxa}, while the future gravitational-wave (GW) astronomy~\cite{LIGOScientific:2016aoc} and black hole (BH) shadow measurements~\cite{EventHorizonTelescope:2019dse} will allow us to clarify gravitational physics in the so-called strong-field regimes as in the vicinity of BHs and neutron stars~\cite{Berti:2015itd,Barack:2018yly,Berti:2018cxi,Berti:2018vdi}.
On the other hand, standard cosmological model based on GR has been plagued by tensions of today's measurements~\cite{Riess:2019cxk,DiValentino:2021izs}, which led to the question for the validity of GR on cosmological distance scales.
In order to solve these tensions, the gravitational theories other than GR have been extensively studied~\cite{Sotiriou:2008rp,Clifton:2011jh,Will:2014kxa,Berti:2015itd}.

One of the simplest and most robust modifications to GR are provided by scalar-tensor (ST) theories which possess a scalar field (denoted by $\phi$) DOF as well as the metric tensor (denoted by $g_{\mu\nu}$) DOFs ~\cite{Fujii:2003pa}.
Traditionally, ST theories which include (non)canonical kinetic terms and/or nonminimal coupling to the spacetime curvature have been applied to inflationary universe and/or dark energy models (see e.g., Refs.~\cite{Sotiriou:2008rp,Clifton:2011jh,Copeland:2006wr,Armendariz-Picon:1999hyi}).
The framework of the ST theories have been extensively generalized by the (re)discovery of the Horndeski theories \cite{Horndeski:1974wa,Deffayet:2009wt,Kobayashi:2011nu}, which are known as the most general ST theories with second-order equations of motion, despite the existence of higher-derivative interactions of the scalar field $\phi$ and the nonminimal derivative coupling to the spacetime curvature. 
The Horndeski theories are characterized by the four independent coupling functions $G_{2,3,4,5} (\phi,X)$, where $X:=-(1/2)g^{\mu\nu}\nabla_\mu\phi\nabla_\nu\phi$ represents the canonical kinetic term of the scalar field with $\nabla_\mu$ being the covariant derivative associated with the metric $g_{\mu\nu}$.
The framework of the Horndeski theories has been extended to the Degenerate-Higher-Order-Scalar-Tensor (DHOST) theories \cite{Langlois:2015cwa,BenAchour:2016fzp} and beyond-DHOST theories \cite{Takahashi:2021ttd,Takahashi:2022mew,Naruko:2022vuh,Takahashi:2023vva}, which eliminate the Ostrogradski ghosts by imposing the degeneracy conditions among the higher-derivative equations of motion.
The existence of BH solutions and their properties will be very important in distinguishing such new class of ST theories from the theoretical perspectives.
This offers an interesting possibility for probing the possible deviation from GR in strong field regimes.

In GR, the uniqueness theorem states that an asymptotically flat, stationary and axisymmetric BH is described by the Kerr solution, which is characterized only by mass and angular momentum~\cite{Israel:1967wq,Carter:1971zc,Ruffini:1971bza}.
This is reduced to the Schwarzschild solution in the limit of static and spherically symmetric spacetime.
The BH no-hair theorem, which states that only the BH solutions are Schwarzschild or Kerr solutions in the case of vacuum spacetime.
We can extend the theorem to the case with a scalar field, assuming an appropriate condition on the potential.
It also holds for the various ST theories with a canonical scalar field~$\phi$~\cite{Hawking:1971vc,Bekenstein:1972ny}, a generalized kinetic term~\cite{Graham:2014mda}, as well as a scalar field nonminimally coupled to the scalar curvature~$F(\phi)R$~\cite{Hawking:1972qk,Bekenstein:1995un,Sotiriou:2011dz,Faraoni:2017ock}.
In the shift-symmetric Horndeski theories which are invariant under the constant shift transformation~$\phi \to \phi+c$, where the functions~$G_{2,3,4,5}$ depend only on $X$, Ref.~\cite{Hui:2012qt} showed that a no-hair result of static and spherically symmetric BH solutions holds under the following hypotheses: 
(i) the scalar field shares the same symmetry as the static and spherically symmetric metric;
(ii) the spacetime is asymptotically flat with a vanishing radial derivative~$\psi'(r) \to 0$ at spatial infinity ($r \to \infty$);
(iii) the norm of the Noether current associated with the shift symmetry $J_\mu J^\mu$ is finite on the BH event horizon;
(iv) a canonical kinetic term~$X$ is present in the Lagrangian;
(v) the $X$-derivatives of $G_{2,3,4,5}$ contain only positive or zero powers of $X$. 
If we violate at least one of the conditions given above, it is possible to realize hairy BH solutions endowed with nontrivial scalar hair.
The no-hair theorem for the static and spherically symmetric BH solutions has been extended to the case of the shift-symmetric beyond-Horndeski theories in Ref.~\cite{Babichev:2016rlq}.
The no-hair theorem in the shift-symmetric Horndeski theories for BH solutions has been generalized to the case of the stationary and axisymmetric BHs in Ref.~\cite{Capuano:2023yyh}.

For a scalar field with the linear dependence on time~$t$ of the form $\phi=qt+\psi(r)$ with $q$ being constant, which evades the hypothesis~(i),  there exist the stealth Schwarzschild solution~\cite{Babichev:2013cya,Kobayashi:2014eva,Babichev:2015rva} and the BH solutions with asymptotically (anti-)de Sitter~[(A)dS] spacetimes \cite{Babichev:2013cya,Babichev:2016fbg}.
If the asymptotic flatness of spacetime is not imposed, which evades the hypothesis~(ii), the linear quartic derivative coupling $X$ in $G_4$ gives rise to the exact hairy BH solutions with an asymptotic geometry mimicking the Schwarzschild--AdS spacetime~\cite{Rinaldi:2012vy,Anabalon:2013oea,Minamitsuji:2013ura,Cisterna:2014nua}.
For the coupling $G_5\propto \ln |X|$, which is equivalent to the linear coupling to the Gauss-Bonnet (GB) term $\phi R_{\rm GB}^2$~\cite{KYY}, where 
\be
\label{gb}
R_{\rm GB}^2:= R^2-4R_{\mu\nu}^2+R_{\mu\nu\alpha\beta}^2,
\ee
 is the GB term, there exists the asymptotically-flat hairy BH solution whose metric components are corrected by the GB coupling~\cite{Sotiriou:2013qea,Sotiriou:2014pfa}. 
There also exists asymptotically-flat BH solution in the model where $G_4(X)\supset (-X)^{1/2}$~\cite{Babichev:2017guv}.
These solutions arise from the violation of the hypothesis~(v).
We note that there also exist the hairy BH solutions for non-shift-symmetric GB couplings $e^{-c\phi} R_{\rm GB}^2$ with $c$ being constant~\cite{Kanti:1995vq,Kanti:1997br,Torii:1996yi} and for BH scalarization models which occur for $Z_2$ symmetric coupling functions~\cite{Silva:2017uqg,Doneva:2017bvd,Antoniou:2017acq,Blazquez-Salcedo:2018jnn,Minamitsuji:2018xde,Silva:2018qhn,Cunha:2019dwb,Konoplya:2019fpy,Doneva:2021dqn,East:2021bqk,Julie:2022huo,Doneva:2020nbb,Doneva:2022yqu,Dima:2020yac,Herdeiro:2020wei,Lai:2023gwe}.

The linear stability analysis of the static and spherically symmetric BH solutions in the Horndeski theories have been performed in the literature, e.g., \cite{Kobayashi:2012kh,Kobayashi:2014wsa,Kase:2021mix}.
These linear stability conditions have been applied to various static and spherically symmetric BH solutions with the nontrivial profile of the scalar field in the Horndeski theories in Refs. \cite{Minamitsuji:2022mlv,Minamitsuji:2022vbi}. 
In generic Horndeski theories, static and spherically symmetric BH solutions with a non-vanishing constant kinetic term on the horizon $X\neq 0$ inevitably suffer from a ghost or gradient instability \cite{Minamitsuji:2022vbi} including the solutions discussed in Refs.~\cite{Rinaldi:2012vy,Anabalon:2013oea,Minamitsuji:2013ura,Cisterna:2014nua}.
On the other hand, it was shown that within the perturbative regime the static and spherically symmetric BH solutions in ST theories with the power-law couplings to the GB invariant are free from the ghost or gradient instability, which include asymptotically-flat BH solutions in the shift-symmetric theory with the linear coupling to the GB invariant $G_5(X) \propto\ln |X|$~\cite{Sotiriou:2013qea,Sotiriou:2014pfa}.
However, in the models where the scalar field linearly depends on time, e.g., the stealth Schwarzschild solution \cite{Babichev:2013cya,Babichev:2016fbg}, the standard linear perturbation analysis cannot be applied because the perturbations become infinitely strongly coupled \cite{deRham:2019gha,Motohashi:2019ymr}.
Thus, we will apply another way to see BH stability, that is BH thermodynamics.
When we have two BH solutions, we can compare their entropies and then argue that the BH solution with smaller entropy is more unstable than the other BH solution with larger entropy. 
In this paper, we will focus on BH thermodynamics in the Horndeski theories.
Although the Wald entropy formula \cite{Wald:1993nt} has been useful for computing the BH entropy in the covariant gravitational theories which contain the dependence on the Riemann tensor, this may not be directly applicable to the Horndeski theories because of the presence of the derivative interactions of the scalar field and the nonminimal derivative couplings of the scalar field to the spacetime curvature tensors \cite{Feng:2015oea,Hajian:2020dcq}.
The terms which contain the spacetime curvature tensors may be replaced with the higher-derivatives of the scalar field with use of the properties of the Riemann tensor and the partial integration of the action.
The apparent dependence of the action on the spacetime curvature tensors may be modified before and after a partial integration, although the action after the partial integration is equivalent to that before the partial integration under the assumption of no contribution from the boundaries.
In this work, following the original formulation by Iyer and Wald~\cite{Iyer:1994ys}, we will construct the BH thermodynamics in the Horndeski theories from the first principle.
Since the Horndeski theories preserve the four-dimensional diffeomporphism invariance, there exists the associated Noether charge potential whose explicit form was obtained in Ref.~\cite{Peng:2015yjx}.
Since the Noether charge potential is independent of the apparent modification of the action by the partial integration, this should be able to provide the unique description of the first-law of BH thermodynamics and the BH entropy.
Iyer and Wald showed that the variation of the Hamiltonian is given by that of the Noether charge potential evaluated on the boundaries, i.e., in our case, the BH event horizon and spatial infinity \cite{Iyer:1994ys}.
The conservation of the total Hamiltonian of the BH system reproduces the first law of the BH thermodynamics.
Our theorem will be able to be applied to both the shift-symmetric and non-shift-symmetric subclass of the Horndeski theories.
Previously, for a subclass of the Horndeski theories with the nonminimal coupling to the Einstein tensor $G^{\mu\nu}\nabla_\mu\phi \nabla_\nu \phi$, the Iyer-Wald formulation has been applied for the BH solutions without the electric field in Refs. \cite{Feng:2015oea,Hajian:2020dcq}
and with the electric field in Ref.~\cite{Feng:2015wvb}.
The Iyer-Wald formulation has also been applied to the planar BH solutions in some classes of the Horndeski theories in arbitrary dimensions~\cite{Bravo-Gaete:2021hlc}.
Our analysis will cover the whole Horndeski theories and be able to apply all the static and spherically symmetric BH solutions including those with the linearly time-dependent scalar field in the shift-symmetric theories \cite{Babichev:2013cya,Babichev:2016fbg}.

The paper is constructed as follows:
In Sec. \ref{sec2}, we apply the formulation by Iyer and Wald to the Horndeski theories. In Sec. \ref{sec3}, we discuss the entropy and mass for the static and spherically symmetric BH solutions with the static scalar field in the Horndeski theories. 
In Sec. \ref{sec4}, we discuss the entropy and mass of the system for the BH solutions with the linearly time-dependent scalar field in the shift-symmetric Horndeski theories. 
In Sec. \ref{sec5}, we investigate thermodynamical stability of the stealth Schwarzschild BH solutions and the Schwarzschild-(A)dS BH solutions with the linear time dependence in the shift-symmetric Horndeski theories which are discussed in Sec. \ref{sec4}. 
In Sec. \ref{sec6},~we consider the Horndeski theories minimally coupled to the $U(1)$-invariant vector field where BH solutions contain the mass and the electric charge, and clarify the conditions under which the differential of the BH entropy is integrable in spite of the presence of the two independent charges.
The last Sec. \ref{sec7} is devoted to giving a brief summary and conclusion.

\section{Iyer-Wald formulation in the Horndeski theories}
\label{sec2}

\subsection{The Horndeski theories}

We consider the Horndeski theories \cite{Horndeski:1974wa,Deffayet:2009wt,Kobayashi:2011nu}
whose action is composed of the four independent parts
\be
\label{action}
S
&=&
\int d^4x\sqrt{-g}
 {\cal L}
=
\int d^4x\sqrt{-g}
\sum_{i=2}^5 {\cal L}_i,
\ee
with the Lagrangian densities given by 
\be
{\cal L}_2
&:=&
G_2(\phi, X),
\label{l2}
\\
{\cal L}_3
&:=&
-G_3(\phi,X)\Box\phi,
\label{l3}
\\
{\cal L}_4
&:=&
G_4(\phi, X) R
+G_{4X}(\phi, X)
\left[
\big(\Box\phi\big)^2
-
\left(
\phi^{\alpha\beta}
\phi_{\alpha\beta}
\right)
\right],
\label{l4}
\\
{\cal L}_5
&
:=&
G_5(\phi, X) G_{\mu\nu}\phi^{\mu\nu}
-
\frac{1}{6}G_{5X}(\phi, X)
\left[
(\Box\phi)^3
-3\Box \phi
\left(
\phi^{\alpha\beta}
\phi_{\alpha\beta}
\right)
+
2
\phi_{\alpha}{}^\beta
\phi_\beta{}^\rho
\phi_\rho{}^\alpha
\right],
\label{l5}
\ee
where $g_{\mu\nu}$ is the spacetime metric, $R$ and $G_{\mu\nu}$ are the Ricci scalar and Einstein tensor associated with the metric $g_{\mu\nu}$, respectively, $\phi$ is the scalar field, $\phi_\mu=\nabla_\mu \phi$, $\phi_{\mu\nu}=\nabla_\mu\nabla_\nu\phi$, and so on are the short-hand notation for the covariant derivatives of the scalar field, with $\nabla_\mu$ being the covariant derivative associated with the metric $g_{\mu\nu}$. 
$X$ represents the canonical kinetic term $X:=-(1/2)g^\mn\phi_\mu \phi_\nu$ with use of the short-hand notation, and $G_{2,3,4,5}(\phi,X)$ are the free functions of $\phi$ and $X$.
We also define $\phi^\mu{}_\nu:=g^{\mu\alpha}\phi_{\alpha\nu}$, $\phi^{\mu\nu}:=g^{\nu\alpha}\phi^\mu{}_\alpha$, and $\Box \phi:= g^{\mu\nu}\phi_{\mu\nu}$.

\subsection{The Noether charge potential associated with the diffeomorphism invariance}

The variation of the action \eqref{action} is given by 
\be
\delta S
=
\int d^4x
\sqrt{-g}
\left(
E_{\mu\nu} 
\delta g^{\mu\nu}
+
E_\phi
\delta \phi
+
\nabla_\mu J^\mu
\right),
\ee
where the equations of motion of the metric and scalar field, $E_{\mu\nu}=0$ and $E_\phi=0$, respectively, can be found in Ref. \cite{Kobayashi:2011nu} for instance, and the boundary current is given by~\cite{Peng:2015yjx} 
\be
J^\mu=\sum_{i=2}^5J_{(i)}^\mu,
\ee
which is composed of the parts from the Lagrangians \eqref{l2}-\eqref{l5}
\be
\label{j2}
J_{(2)}^\mu
&=&-G_{2X} \phi^\mu \delta\phi,
\\
\label{j3}
J_{(3)}^\mu
&=&
-\frac{1}{2}
G_3
\left(
\mh \phi^\mu
-2\mh^{\mu\nu}\phi_\nu
+2\nabla^\mu\delta\phi
\right)
+
\delta\phi 
G_{3X}
\Box\phi
\phi^\mu
+
\delta\phi
\nabla^\mu 
G_3,
\\
\label{j4}
J_{(4)}^\mu
&=&
G_{4X} \Box \phi
\left(
\mh \phi^\mu-2\mh^{\mu\nu}\phi_\nu
\right)
+
2G_{4X}
\Box\phi 
\nabla^\mu (\delta\phi)
-
2\nabla^\mu
\left(
G_{4X}\Box\phi
\right)
\delta\phi
\nonumber
\\
&&
+
G_{4XX}
\left(
\phi^{\alpha\beta}\phi_{\alpha\beta}
-(\Box\phi)^2
\right)
\phi^\mu \delta\phi
+
G_{4X}
\left(
2\phi^{\mu\rho}\phi^\sigma
-
\phi^{\rho\sigma}\phi^\mu
\right)
\mh_{\rho\sigma}
-
2G_{4X}
\phi^{\mu\nu}\nabla_\nu\delta\phi
\nonumber\\
&&
+
2\nabla_\nu\left(G_{4X}\phi^{\mu\nu}\right)
\delta\phi
-
\mh^{\mu\nu}
\nabla_\nu G_4
+
G_4
\nabla_\nu \mh^{\mu\nu}
+
\mh
\nabla^\mu G_4
-
G_4
\nabla^\mu \mh
-
G_{4X}
R\phi^\mu\delta\phi,
\\
\label{j5}
J_{(5)}^\mu
&=&
\frac{1}{4}G_{5X}
\phi^{\alpha\beta}\phi_{\alpha\beta}
\left(
   \mh \phi^\mu
-2\mh^{\mu\nu}\phi_\nu
\right)
-
\frac{1}{2}
\mh_{\rho\sigma}
G_{5X}
\Box\phi
\left(
2\phi^{\sigma\mu}\phi^\rho
-
\phi^{\sigma\rho}\phi^\mu
\right)
\nonumber
\\
&&
-
\frac{1}{4}
G_{5X}
(\Box\phi)^2
\left(
\mh \phi^\mu
+
2\nabla^\mu \delta\phi
-
2\mh^{\mu\nu}
\phi_\nu
\right)
+\frac{1}{2}
\nabla^\mu
\left[
G_{5X}
(\Box\phi)^2
\right]
\delta\phi
\nonumber\\
&&
+
\frac{1}{2}
G_{5X}
\phi^{\alpha\beta}\phi_{\alpha\beta}
\nabla^\mu\delta\phi
-
\frac{1}{2}
\delta\phi
\nabla^\mu
\left[
G_{5X}
\left(\phi_{\alpha\beta}\right)^2
\right]
+
G_{5X}
\Box\phi
\phi^{\mu\nu}
\nabla_\nu\delta\phi
\nonumber\\
&&
-\delta\phi
\nabla_\nu
\left(
G_{5X}
\Box\phi
\phi^{\mu\nu}
\right)
-
G_{5X}
\phi^{\mu\sigma}
\phi_{\nu\sigma}
\nabla^\nu
\delta\phi
+
\delta\phi
\nabla^\nu
\left(
G_{5X} \phi^{\mu\sigma}\phi_{\nu\sigma}
\right)
\nonumber\\
&&
+
\frac{1}{6}
G_{5XX}
\left[
(\Box\phi)^3
-3\Box \phi
(\phi_{\alpha\beta})^2
+
2
(\phi_{\alpha\beta})^3
\right]
\phi^\mu \delta\phi
+
\frac{1}{2}
G_{5X}
\phi^\nu_\sigma
\left(
2\phi^{\sigma\mu}\phi^\rho
-\phi^{\sigma\rho} \phi^\mu
\right)
\mh_{\rho \nu}
\nonumber
\\
&&
-
\mh_{\rho\sigma}
\nabla^\sigma
\left(
G_5 \phi^{\mu\rho}
\right)
+
G_{5}
\phi_{\sigma\rho}\nabla^{\sigma}
\mh^{\rho\mu}
-
\frac{1}{2}
G_5 
\phi_{\rho\sigma}
\nabla^{\mu}
\mh^{\rho\sigma}
+\frac{1}{2}
\mh^{\rho\sigma}
\nabla^\mu
\left(
G_5 \phi_{\rho\sigma}
\right)
-
\frac{1}{2}
G_5 
\phi^{\mu\nu}
\nabla_\nu \mh
+
\frac{1}{2} 
\mh
\nabla_\nu
\left(
G_5 \phi^{\mu\nu}
\right)
\nonumber\\
&&
-
\frac{1}{2}
G_5
\Box\phi
\nabla_\rho \mh^{\rho\mu}
+
\frac{1}{2}
\mh^{\rho\mu}
\nabla_\rho
\left(
G_5 \Box\phi
\right)
+
\frac{1}{2}
G_5
\Box\phi
\nabla^\mu 
\mh
-\frac{1}{2}
\mh
\nabla^\mu
\left(
G_5\Box\phi
\right)
-
G_5 
\mh_{\rho\sigma}
G^{\mu\rho}
\phi^\sigma
+
\frac{1}{2}
G_5
\mh_{\rho\sigma}
G^{\rho\sigma}
\phi^\mu
\nonumber
\\
&&
-\delta \phi
G^{\mu\nu}
\nabla_\nu 
G_5
+
G_5
G^{\mu\nu}
\nabla_\nu \delta \phi
-
\delta \phi
G_{5X}
G^{\rho\sigma}
\phi_{\rho\sigma}
\phi^\mu,
\ee
where we have defined the variation of the metric tensor with respect to the independent integration constants by 
\be
\mh_{\mu\nu}=\delta g_{\mu\nu},
\qquad
\mh^{\mu\nu}=g^{\mu\rho}g^{\nu\sigma}
\mathfrak{h}_{\rho\sigma},
\qquad 
\mh=g^{\rho\sigma} \mathfrak{h}_{\rho\sigma}.
\ee
We also define the dual 3-form to $J^\mu$ by 
\be
\label{dual_3form}
{\Theta}_{\alpha\beta\gamma}
&:=&
J^\mu
\varepsilon_{\mu\alpha\beta\gamma}
=
{\varepsilon}_{\alpha\beta\mu \gamma}
J^\mu
=
\sum_{i=2}^5
{\varepsilon}_{\alpha\beta\mu \gamma}
J_{(i)}^\mu.
\ee
Since the Horndeski theories~\eqref{action} with Eqs. \eqref{l2}-\eqref{l5} are invariant under the four-dimensional diffeomorphism transformation, $x^\mu\to x^\mu+\xi^\mu (x^\mu)$, there exists the associated Noether charge potential. 
Under the diffeomorphism transformation, the variations of the metric and scalar field are, respectively, given by 
\be
\label{diffeo}
\mathfrak{h}^{(\xi)}_{\mu\nu}
&=&
\delta_\xi g_{\mu\nu}
:=
\mathcal{L}_{\xi}
g_{\mu\nu}
=2 \nabla_{(\mu}\xi_{\nu)},
\qquad
\delta_\xi \phi
:=
\mathcal{L}_{\xi}\phi
=
\xi^\mu \phi_\mu,
\ee
and with use of the on-shell gravitational equations of motion $E_{\mu\nu}=0$, $J_{(\xi)}^\mu$ can be written in terms of the total derivative of the Noether charge potential $K^{\mu\nu}_{(\xi)}$, i.e., 
\be
\label{jmu}
J_{(\xi)}^\mu 
-
\xi^{\mu} 
{\cal L}
=
2
\nabla_\nu K^{[\nu\mu]}_{(\xi)}
=
2
\sum_{i=2}^5
\nabla_\nu
 K^{[\nu\mu]}_{(i)(\xi)},
\ee
where each individual contribution
is given by 
\be
\label{k2}
K^{\mu\nu}_{(2)(\xi)}
&=&
0,
\\
\label{k3}
K^{\mu\nu}_{(3)(\xi)}
&=&
-
G_3 \xi^{\mu}\phi^{\nu},
\\
\label{k4}
K^{\mu\nu}_{(4)(\xi)}
&=&
2G_{4X}
\Big[
\Box\phi \xi^{\mu}\phi^{\nu}
-\xi_\sigma 
\phi^{\sigma\mu}
\phi^{\nu}
\Big]
+2
\xi^{\mu}\nabla^{\nu}G_4
+
G_4 
\nabla^{\mu}\xi^{\nu},
\\
\label{k5}
K^{\mu\nu}_{(5)(\xi)}
&=&
-
\frac{1}{2}
G_{5X}
\left[
\left(
\Box\phi^2
-
\phi^{\alpha\beta}
\phi_{\alpha\beta}
\right)
\xi^{\mu}\phi^{\nu}
+
2
\left(
\xi^\rho \phi_{\rho\sigma}
-\Box\phi \xi_\sigma
\right)
\phi^{\sigma\mu}
\phi^{\nu}
\right]
\nonumber\\
&&
+
\xi^{\mu} \nabla_\sigma \left( \phi^{\nu\sigma}G_5\right)
-
\xi_\sigma
\nabla^{\mu}
\left(
\phi^{\nu\sigma}
G_5
\right)
-
\xi^{\mu}
\nabla^{\nu}
\left(
G_5\Box\phi
\right)
+
\frac{1}{2}
G_5
\left(
2\xi_\sigma G^{\sigma\mu}\phi^{\nu}
-2(\nabla_\sigma \xi^{\mu})\phi^{\nu\sigma}
-\Box\phi
  \nabla^{\mu}\xi^{\nu}
\right).
\ee
We then define the dual 2-form of the Noether charge potential $K_{(\xi)}^{\mu\nu}$ \cite{Peng:2015yjx}
\be
\label{def_q}
{Q}_{(\xi)\alpha\beta}
:=
-\epsilon_{\alpha\beta\mu\nu}
K^{\mu\nu}_{(\xi)}
=
\sum_{i=2}^5
{Q}^{(i)}_{(\xi)\alpha\beta}.
\ee
We also define the 2-form tensor where the first index of ${\Theta}_{\nu\alpha\beta}$ defined in Eq.~\eqref{dual_3form} is contracted by the infinitesimal differmorphism transformation $\xi^\nu$, by 
\be
\label{itheta}
i_\xi {\Theta}_{\alpha\beta}
&:=&
\xi^\nu
{\Theta}_{\nu\alpha\beta}
=
-
{\varepsilon}_{\alpha\mu \beta\nu}
J^\mu\xi^\nu
=
\varepsilon_{\alpha\beta\mu\nu}
J^\mu \xi^\nu.
\ee
We now consider the variation of the dual Noether charge potential with respect to the physical parameters
subtracted by Eq.~\eqref{itheta}
\be
\label{deltaQ}
\delta{Q}_{(\xi)\alpha\beta}
-
i_\xi {\Theta}_{\alpha\beta}
=
-
\left(
\delta
\left(
\varepsilon_{\alpha\beta\mu\nu}
K^{\mu\nu}_{(\xi)}
\right)
+
\varepsilon_{\alpha\beta\mu\nu}
J^\mu \xi^\nu
\right)
=
-
\sum_{i=2}^5
\left(
\delta
\left(
\varepsilon_{\alpha\beta\mu\nu}
K^{\mu\nu}_{(i)(\xi)}
\right)
+
\varepsilon_{\alpha\beta\mu\nu}
J_{(i)}^\mu \xi^\nu
\right).
\ee
The integration of Eq.~\eqref{deltaQ} on the boundaries of the Cauchy surface gives rise to the variation of the Hamiltonian \cite{Wald:1993nt,Iyer:1994ys}.

\subsection{Static and spherically symmetric black hole solutions}

We consider the static and spherically symmetric solutions whose metric is written by 
\be
\label{metric}
ds^2
&=&
-h(r) dt^2
+\frac{dr^2}
          {f(r)}
+r^2\gamma_{ab}d\theta^a d\theta^b,
\ee
where $t$ and $r$ are the temporal and radial coordinates, and $\gamma_{ab}d\theta^a d\theta^b :=d\theta^2+\sin^2\theta d\varphi^2$ represents the metric of the unit two-sphere.
We assume that the spacetime contains the event horizon at $r=r_g$ where
\be
\label{horizon_condition}
h(r_g)=f (r_g)=0,
\qquad 
\lim_{r\to r_g}
\frac{f(r)}{h(r)}
=
{\rm const}.
\ee
In the case that $h(r)$ and $f(r)$ have several roots, we assume that $r_g$ corresponds to the largest positive root, and in the entire domain outside the event horizon $r_g<r< \infty$ the two metric functions $f(r)$ and $h(r)$ are regular and positive.
For the scalar field, we will consider the two ansatze; the static ansatz \eqref{static_sc} or the ansatz with the linear time dependence \eqref{linear_time_sc}, where the latter can be applied only for the shift-symmetric Horndeski theories.
In this background \eqref{metric}, we assume that $\xi^\mu$ corresponds to the timelike Killing vector field, $\xi^\mu=(1,0,0,0)$.

The variations of the metric and scalar field  of a given BH solution can be written in terms of those of the integration constants
\be
\label{variation}
&&
\mathfrak{h}_{tt}
=
-\delta h
=
-\sum_j \frac{\partial h}{\partial c_j}\delta c_j,
\quad 
\mathfrak{h}_{rr}
=-\frac{\delta f }{f^2}
=
-\frac{1}{f^2}
\sum_j
\frac{\partial f}{\partial c_j}
\delta c_j,
\quad 
\mathfrak{h}_{ab}
=0,
\quad 
\delta \phi
=
\sum_j
\frac{\partial \phi}{\partial c_j}
\delta c_j,
\ee
where $c_j$'s are integration constants of the BH solutions, for instance, the position of the event horizon $r_g$.

As shown in Refs.~\cite{Wald:1993nt,Iyer:1994ys}, with use of Eq.~\eqref{deltaQ}, the variation of the Hamiltonian with respect to the integration constants in a specific solution is given by the contributions from the boundaries, i.e., the horizon $r\to r_g$ and infinity $r\to \infty$,
\be
\label{hamiltonian}
\delta {\cal H}
&:=&
\delta{\cal H}_\infty
-
\delta{\cal H}_{H}
\nonumber
\\
&=&
-
\int 
d\Omega
\left(
\delta
\left(
r^2 \sqrt{\frac{h}{f}}
K^{[tr]}_{(\xi)}
\right)
+
r^2 \sqrt{\frac{h}{f}}
J^{[t} \xi^{r]}
\right)
\Big|_{r\to\infty}
+
\int 
d\Omega
\left(
\delta
\left(
r^2 \sqrt{\frac{h}{f}}
K^{[tr]}_{(\xi)}
\right)
+
r^2 \sqrt{\frac{h}{f}}
J^{[t} \xi^{r]}
\right)
\Big|_{r\to r_g}
\nonumber\\
&=&
-
\int
d\Omega
\sum_{i=2}^5
\left(
\delta
\left(
r^2
\sqrt{\frac{h}{f}}
K^{[tr]}_{(i)(\xi)}
\right)
+
r^2
\sqrt{\frac{h}{f}}
J_{(i)}^{[t} \xi^{r]}
\right)
\Big|_{r\to\infty}
\nonumber
\\
&&+
\int
d\Omega
\sum_{i=2}^5
\left(
\delta
\left(
r^2
\sqrt{\frac{h}{f}}
K^{[tr]}_{(i)(\xi)}
\right)
+
r^2
\sqrt{\frac{h}{f}}
J_{(i)}^{[t} \xi^{r]}
\right)
\Big|_{r\to r_g},
\ee
where 
$d\Omega:=\sin\theta d\theta d\varphi$ and the subscript `$H$' represents the quantities associated with the horizon.
The variation of the Hamiltonian on the horizon and at the infinity can be identified with the variation of the total mass of the system $M_{\rm H}$ and BH entropy $S_{\rm H}$ in the Horndeski theories as
\be
\label{var_hamilton}
\delta {\cal H}_\infty=\delta M_{\rm H},
\qquad 
\delta {\cal H}_{H}
=T_{\textsf{H}({\rm H})} 
\delta S_{\rm H},
\ee
where $T_{\textsf{H}({\rm H})}$ represents the Hawking temperature of the given BH solution
\be
T_{\textsf{H}({\rm H})}:=\frac{\sqrt{h'(r_g)f'(r_g)}}{4\pi}.
\label{hawking_temperature}
\ee
The conservation of the total Hamiltonian, $\delta {\cal H}=0$, reproduces the first law of the BH thermodynamics in the Horndeski theories
\be
T_{\textsf{H}({\rm H})} \delta S_{\rm H}=\delta M_{\rm H}.
\ee
We note that in some classes of the Horndeski theories GWs may propagate with the speeds different from the speed of light.
In such a case, there was the argument that the Hawking temperature should be evaluated on the horizon of the effective metric for GWs, which are disformally related to the original metric $g_{\mu\nu}$ \cite{Hajian:2020dcq}.
Here, we choose the surface gravity for the original metric $g_{\mu\nu}$
as the Hawking temperature \eqref{hawking_temperature} as in the case of GR.
The first reason is because photons and other massless particles as the products of the Hawking evaporation would propagate along the light cones of the original metric $g_{\mu\nu}$.
The second reason is because there is no unique choice of the frames where GWs travel with the speed of light.
Especially, the conformal transformation does not modify the speeds of GWs, but red-shifts or blue-shifts the Hawking temperature.

\subsubsection{The case of the static scalar field}

We now compute the integrand of Eq.~\eqref{hamiltonian}.
First, we focus on the solution with the static scalar field
\be
\phi=\psi(r).
\label{static_sc}
\ee
Under the variation \eqref{variation}
the integrand of the variation of the Hamiltonian \eqref{hamiltonian}
is given by 
\be
\label{integrand__non_qt}
&&
-
\delta
\left(
r^2
\sqrt{\frac{h}{f}}
K^{[tr]}_{(\xi)}
\right)
-
r^2
\sqrt{\frac{h}{f}}
J^{[t}\xi^{r]}
\nonumber\\
&=&
2
r^2
\sqrt{\frac{h}{f}}
\Big\{
-\frac{1}{2}
f\psi' 
\delta\psi
G_{2X}
+
f\psi' \delta\psi
G_{3\phi}
+
\frac{f\psi'^2}{4rh}
\left[
f(4h+rh') \delta\psi
-
rh (2f \delta\psi'+\psi'\delta f)
\right]
G_{3X}
\nonumber\\
&&
-
\frac{\delta f}{r}
G_4
+
\frac{1}{2}
\left[
f
\left(
-2\delta \psi'
+
\frac{h'}{h}
\delta\psi
\right)
-
\psi'\delta f
\right]
G_{4\phi}
+
\frac{f\psi'}{r^2h}
\left[
\left(
(f-1)h+r f h'
\right)
\delta\psi
-
2r h (f\delta\psi'+ \psi' \delta f)
\right]
G_{4X}
\nonumber\\
&&
-f\psi' \delta \psi  G_{4\phi\phi}
+
\frac{f\psi'^2}{2rh}
\left[
-f(8h+rh')\delta\psi
+ rh (2f\delta \psi'+\psi'\delta f)
\right]
G_{4\phi X}
\nonumber\\
&&
+
\frac{f^2\psi'^3}{r^2h}
\left[
-f(h+rh')\delta\psi
+ rh (2f\delta \psi'+\psi'\delta f)
\right]
G_{4X X}
+
\frac{f\psi'}{2r^2h}
\left[
-2
\left(
(f-1)h+rfh'
\right)
\delta\psi
+
rh
\left(
4f\delta\psi'
+
3\psi' \delta f
\right)
\right]
G_{5\phi}
\nonumber
\\
&&
+
\frac{f\psi'^2}{4r^2 h}
\left[
f^2 
(6h\delta \psi'-3h'\delta\psi)
-
h\psi' \delta f
+
f
\left(
h'\delta\psi
+h 
(-2\delta\psi'
+
5\psi'\delta f
)
\right)
\right]
G_{5X}
+
\frac{f^2\psi'^2\delta \psi}
       {r}
G_{5\phi\phi}
\nonumber
\\
&&
+
\frac{f^2\psi'^3}
        {2r^2h}
\left[
f
(2h+rh')
\delta \psi
-
rh
 (2f\delta \psi'+\psi'\delta f)
\right]
G_{5\phi X}
-
\frac{f^3\psi'^4}{4r^2h}
\left[
f(2h\delta\psi'- h'\delta \psi)
+
h\psi' \delta f
\right]
G_{5X X}
\Big\}.
\ee

\subsubsection{The case of the scalar field with linear time dependence}

Second, we consider the shift-symmetric Horndeski theories invariant under the constant shift $\phi\to \phi+c$ with $c$ being constant, which correspond to the theories without the dependence on $\phi$ in the coupling functions:
\be
\label{shift_symmetric_subclass}
G_2=G_2 (X),
\qquad 
G_3=G_3 (X),
\qquad 
G_4=G_4 (X),
\qquad 
G_5=G_5 (X).
\ee
There is the Noether current associated with the shift symmetry
\be
{\cal J}^\mu
=
\frac{1}
       {\sqrt{-g}}
\left[
\frac{\partial  {\cal L}}
        {\partial \phi_\mu}
-
\nabla_\nu
\left(
\frac{\partial  {\cal L}}
        {\partial \phi_{\mu\nu}}
\right)
\right].
\ee
The theory \eqref{shift_symmetric_subclass} admits the static and spherically symmetric BH solutions with the linearly time-dependent scalar field
\cite{Babichev:2013cya,Kobayashi:2014eva,Babichev:2015rva,Babichev:2016fbg}
\footnote{
Because of the linear time dependence, the ansatz for the scalar field \eqref{linear_time_sc} does not respect the symmetry of the spacetime, $\pounds_\xi \phi\neq 0$,
where $\xi^\mu$ corresponds to the timelike Killing vector, while $\pounds_\xi g_{\mu\nu}=0$. However, in deriving the variation of the Hamiltonian \eqref{hamiltonian}, the symmetry $\pounds_\xi \phi= 0$ is not imposed \cite{Iyer:1994ys} and hence our formulation can be applied to the solutions with the scalar field Eq.~\eqref{linear_time_sc}}.
\be
\label{linear_time_sc}
\phi=qt+\psi(r).
\ee
For the metric ansatz \eqref{metric}, the radial component of the Noether current associated with the shift symmetry
is given by 
\be
{\cal J}^r
&=&
-f \psi' G_{2X}
+
\frac{f}
        {2rh^2}
\left[
-q^2 r h'
+
fh 
\left(
4h+rh'
\right)
\psi'^2
\right]
G_{3X}
\nonumber\\
&&
+
\frac{2f \phi'}
        {r^2h}
\left[
(f-1)h
+r f  h'
\right]
G_{4X}
+
\frac{2f^2\psi'}
        {r^2 h^2}
\left[
q^2 rh'
-
fh \left(h+rh'\right)
\psi'^2
\right]
G_{4XX}
\nonumber\\
&+&
\frac{fh'}{2r^2h^2}
\left[
q^2 (f-1)
+
(1-3f)fh\psi'^2
\right]
G_{5X}
+
\frac{f^3 h'\psi'^2}
        {2r^2 h^2}
\left(
-q^2
+
f h \psi'^2
\right)
G_{5XX}.
\ee
For the given ansatz of the metric and scalar field,
Eqs.
\eqref{metric}
and 
\eqref{linear_time_sc},
we can show that the $(t,r)$-component of the metric equations is proportional to ${\cal J}^r$~\cite{Babichev:2013cya,Kobayashi:2014eva,Babichev:2015rva}, and hence we have to impose
\be
\label{jr0}
{\cal J}^r=0.
\ee
The variation of the scalar field is given by
\be
 \delta \phi=
\delta \psi (r).
\ee
We note that since $q$ is not the integration constant but the constant appearing in the ansatz
of the scalar field compatible within the shift symmetry, we do not need to take the variation of $q$ into consideration.
Under the variation \eqref{variation}, the integrand of the variation of the Hamiltonian \eqref{hamiltonian}
is given by 
\be
\label{integrand_qt}
&&
-
\delta
\left(
r^2
\sqrt{\frac{h}{f}}
K^{[tr]}_{(\xi)}
\right)
-
r^2
\sqrt{\frac{h}{f}}
J^{[t}\xi^{r]}
\nonumber\\
&=&
2
r^2
\sqrt{\frac{h}{f}}
\Big\{
-
\frac{f\psi'}{4h^2}
\left[
q^2\delta h
+
h^2
\left(
2 f \psi' \delta\psi'
+\psi'{}^2\delta f
\right)
\right]
G_{3X}
\nonumber\\
&&
-
\frac{\delta f}{r}
G_4
-
\frac{2f \psi'}{r}
\left(
f \delta \psi'
+\psi' \delta f
\right)
G_{4X}
+
\frac{f^2\psi'{}^2}{r h^2}
\left\{
q^2 \delta h
+
h^2
\left(
2f  \psi' \delta \psi'
+\psi'{}^2 \delta f
\right)
\right\}
G_{4XX}
\nonumber\\
&&
+
\frac{f\psi'}{4r^2 h^2}
\left\{
q^2(f-1)\delta h
+
h^2
\big(
6f^2 \psi'\delta\psi'
-
\psi'{}^2 \delta f
+
f
\psi'
(-2\delta\psi'+5\psi'\delta f)
\big)
\right\}
G_{5X}
\nonumber\\
&&
-
\frac{f^3\psi'{}^3}
        {4r^2h^2}
\left\{
q^2\delta h
+
h^2
\left(
2 f \psi' \delta\psi'
+
\psi'{}^2
\delta f
\right)
\right\}
G_{5XX}
+
\frac{1}{2}
\delta \psi
{\cal J}^r
\Big\}.
\ee
We note that with the condition \eqref{jr0}, the terms which are explicitly proportional to $\delta \psi$ vanish.

\section{Black holes with the static scalar field}
\label{sec3}

In this section, we focus on several classes of the Horndeski theories giving rise to the BH solutions with the static scalar field \eqref{static_sc}.

\subsection{GR}

For GR with the cosmological constant $\Lambda$
\be
G_2=-
\frac{1}{8\pi G}
\Lambda ,
\qquad 
G_4=
\frac{1}{16\pi G},
\qquad 
G_3=G_5=0,
\ee
Eq. \eqref{integrand__non_qt} reduces to 
\be
-
\delta
\left(
r^2
\sqrt{\frac{h}{f}}
K^{[tr]}_{(\xi)}
\right)
-
r^2
\sqrt{\frac{h}{f}}
J^{[t}\xi^{r]}
&=&
-
r^2
\sqrt{\frac{h}{f}}
\frac{\delta f}{8\pi Gr}.
\ee
In GR, the Schwarzschild-(A)dS solutions given by 
\be
f(r)=h(r)=1
-\frac{r_g}{3r}
\left(3-r_g^2\Lambda\right)
-\frac{\Lambda}{3}r^2,
\qquad 
\psi(r)=0.
\ee
are the unique static and spherically symmetric BH solution.
Since $r_g$ is only the integration constant, using Eq. \eqref{variation}, $\delta f=\frac{\partial f}{\partial r_g}\delta r_g$ and $\delta h=\frac{\partial h}{\partial r_g}\delta r_g$.
Evaluating Eq.~\eqref{var_hamilton} with use of Eq.~\eqref{hamiltonian}, we obtain the first law of thermodynamics 
\be
T_{\textsf{H}({\rm GR})}
\delta S_{\rm GR}
=
\delta M_{\rm GR}
=
\frac{1}{2G}
\left(
1-r_g^2\Lambda
\right)
\delta r_g,
\ee
where the Hawking temperature \eqref{hawking_temperature} is given by $T_{\textsf{H}({\rm GR})} =T_0(1-r_g^2\Lambda)$.
Here we introduce the Hawking temperature of Schwarzschild BH in GR defined by 
\be
T_0:= \frac{1}{4\pi r_g}.
\label{T0}
\ee
We shall also use the mass and BH entropy of Schwarzschild BH in GR given by
\be
M_0&:=&\frac{r_g}{2G}\,, 
\label{M0}
\\
S_0&:=& \frac{\pi r_g^2}{G}\,, 
\label{S0}
\ee
as reference.

Thus, $\delta S_{\rm GR}=\frac{2 \pi r_g}{G} \delta r_g=\frac{1}{4G} \delta A_{H}$, where $A_{H}:=4\pi r_g^2$ is the area of the BH event horizon, and hence by integrating it we recover the area law
\be
S_{\rm GR}
=
S_0\,,
\label{area_law}
\ee
where we set the integration constant
so that we have the vanishing BH entropy $S_{\rm GR}\to 0$ in the limit of the vanishing horizon radius $r_g\to 0$.
The mass of the system is also given by 
\be
M_{\rm GR}
=
M_0
\left(
1
-
\frac{1}{3}\Lambda r_g^2
\right),
\ee
which coincides with the total mass of the BH, where we set the integration constant so that we have the vanishing mass $M_{\rm GR} \to 0$ in the limit of the vanishing horizon radius $r_g\to 0$.

\subsection{Scalar-tensor theory with nonminimal coupling}

As the next simplest example, we consider the ST theory with nonminimal coupling to the scalar curvature
\be
{\cal L}
=
\omega (\phi)
\left(
R-2V(\phi)
\right)
+\eta X,
\ee
which is equivalent to the Horndeski theory with
\be
G_2
&=&
\eta X
-2\omega (\phi) V(\phi),
\qquad
G_4
=
\omega (\phi),
\qquad
G_3
=
G_5
=
0,
\ee
where $\omega(\phi)$ and $V(\phi)$
are the nonminimal coupling function and the potential of the scalar field, respectively. 
Eq. \eqref{integrand__non_qt} reduces to 
\be
&&
-
\delta
\left(
r^2
\sqrt{\frac{h}{f}}
K^{[tr]}_{(\xi)}
\right)
-
r^2
\sqrt{\frac{h}{f}}
J^{[t}\xi^{r]}
\nonumber
\\
&&
=
2
r^2
\sqrt{\frac{h}{f}}
\Big\{
\left(
-\frac{\omega(\psi)}{r}
-\frac{\psi'}{2}
\omega^{(1)}(\psi)
\right)
\delta f
-
\frac{f}{2 h}
\left[
h'
\omega^{(1)}(\psi)
-
h
\psi'
(\eta
+
2\omega^{(2)}(\psi))
\right]
\delta\psi
-f
\omega^{(1)}(\psi)
\delta\psi'
\Big\},
\ee
where $\omega^{(n)}(\phi)$ and $V^{(n)}(\phi)$ denote the $n(=1,2,\cdots)$-th order derivatives of $\omega(\phi)$ and $V(\phi)$ with respect to $\phi$.
We assume that $V(\phi)$ and $\omega(\phi)$ have their local minima at $\phi=0$, i.e., $V^{(1)}(0)=0$ and $\omega^{(1)}(0)=0$.
Note that even if $V^{(1)}(\phi_0)=0$ and $\omega^{(1)}(\phi_0)=0$ for an arbitrary constant $\phi_0$, we can always make $\phi_0=0$ after a suitable shift of $\phi$.
There is the Schwarzschild-(A)dS solution with the trivial scalar field
\be
f(r)
=
h(r)
=
1
-\frac{r_g}{3r}
\left(3-r_g^2V(0)\right)
-\frac{V(0)}{3}r^2,
\qquad 
\psi(r)=0.
\ee
Since $r_g$ is only the integration constant, using Eq. \eqref{variation}, $\delta f=\frac{\partial f}{\partial r_g}\delta r_g$ and  $\delta h=\frac{\partial h}{\partial r_g}\delta r_g$.
Evaluating Eq.~\eqref{var_hamilton} with use of Eq.~\eqref{hamiltonian}, we obtain the first law of BH thermodynamics
\be
T_{\textsf{H} ({\rm H})}
\delta S_{\rm H}
&=&
\delta M_{\rm H}
=
\frac{1}{2G}
\left(
1
-r_g^2V(0)
\right)
\delta r_g,
\ee
where $\omega(0)=1/(16\pi G)$ with $G$ being the gravitational constant, and the Hawking temperature \eqref{hawking_temperature} is given by $T_{\textsf{H} ({\rm H})}=T_0(1-r_g^2V(0))$.
Thus, $\delta S_{\rm H}=\frac{2 \pi r_g}{G}\delta r_g=\frac{1}{4G} \delta A_{H}$,where $A_{H}=4\pi r_g^2$ is the area of the BH event horizon, and hence by integrating it we recover the area law \eqref{area_law}.
The mass of the system is given by 
\be
M_{\rm H} =
M_0
\left(
1-\frac{1}{3}V(0) r_g^2
\right),
\ee
which coincides with the mass of the BH.

\subsection{The Einstein scalar-Gauss-Bonnet theory}

As one of the nontrivial examples, we consider the Einstein-scalar-GB (EsGB) theory 
\be
\label{sgb}
{\cal L}
=
\frac{1}{16\pi G}
 R
+\eta X
+
k(\phi)
\left(
R^2
-4R^{\alpha\beta}R_{\alpha\beta}
+ R^{\alpha\beta\mu\nu}R_{\alpha\beta\mu\nu}
\right),
\ee
which is equivalent to the class of the Horndeski theories with 
\be
\label{sgb_hor}
G_2
&=&
\eta X
+
8
k^{(4)}(\phi)
X^2
\left(
3-\ln X
\right),
\quad 
G_3
=
4 k^{(3)}(\phi)
X
\left(
7-3\ln X
\right),
\nonumber
\\
G_4
&=&
\frac{1}{16\pi G}
+
4
k^{(2)}(\phi)
X
(2-\ln X),
\quad
G_5
=
-4
k^{(1)}(\phi)
\ln X,
\ee
where $k(\phi)$ is the coupling function, and $k^{(n)}(\phi)$ denotes the $n~(=1,2,\cdots)$-th order derivative of $k(\phi)$ with respect to $\phi$.
This theory has been applied, for instance, to the models of spontaneous scalarization of BHs \cite{Silva:2017uqg,Doneva:2017bvd,Antoniou:2017acq,Blazquez-Salcedo:2018jnn,Minamitsuji:2018xde,Silva:2018qhn,Cunha:2019dwb,Konoplya:2019fpy,Doneva:2021dqn,East:2021bqk,Julie:2022huo,Doneva:2020nbb,Doneva:2022yqu,Dima:2020yac,Herdeiro:2020wei,Lai:2023gwe}.
Eq. \eqref{integrand__non_qt} reduces to
\be
\label{ktr_gb}
&&
-
\delta
\left(
r^2
\sqrt{\frac{h}{f}}
K^{[tr]}_{(\xi)}
\right)
-
r^2
\sqrt{\frac{h}{f}}
J^{[t}\xi^{r]}
\nonumber\\
&=&
r^2
\sqrt{\frac{h}{f}}
\Big\{
-\frac{r
 +
 32\pi G
 (1-3f)k^{(1)}(\psi)\psi'}
          {8\pi G r^2}
\delta f
\nonumber\\
&&
-
\frac{f}{r^2 h}
\left[
4(f-1)h' k^{(1)}(\psi)
+
h \psi'
\left(
r^2\eta-8 (f-1)k^{(2)}(\psi)
\right)
\right]
\delta\psi
+
\frac{8f(f-1)}{r^2}
k^{(1)}(\psi)
\delta\psi'
\Big\}.
\label{linear_gb}
\ee
In the case that the scalar field is regular at the event horizon $r=r_g$
and the solutions can be expanded in the vicinity of $r=r_g$ as 
\be
h(r)&=& h_1(r_g)\left(r-r_g\right)+h_2(r_g) \left(r-r_g\right)^2+{\cal O} \left((r-r_g)^3\right),
\label{h_exp}
\\
f(r)&=& f_1(r_g)\left(r-r_g\right)+f_2(r_g) \left(r-r_g\right)^2+{\cal O} \left((r-r_g)^3\right),
\label{f_exp}
\\
\psi(r)&=& \psi_{H}(r_g)+\psi_1(r_g)\left(r-r_g\right)+\psi_2(r_g) \left(r-r_g\right)^2+{\cal O} \left((r-r_g)^3\right),
\label{phi_exp}
\ee
where the coefficients $h_i(r_g)$, $f_i(r_g)$, and $\psi_i  (r_g)$ ($i=1,2,3\cdots$) are in general functions of $r_g$, and $\psi_{H}(r_g)$ represents the amplitude at the horizon which is also a function of $r_g$, on the horizon $r=r_g$ Eq.~\eqref{ktr_gb} reduces to
\be
\left(-
\delta
\left(
r^2
\sqrt{\frac{h}{f}}
K^{[tr]}_{(\xi)}
\right)
-
r^2
\sqrt{\frac{h}{f}}
J^{[t}\xi^{r]}
\right)_{r\to r_g}
=
\frac{\sqrt{f_1(r_g) h_1(r_g)}}
{8\pi G}
\left(
r_g
+
32
\pi G k^{(1)}[\psi_{H} (r_g)]
\frac{\partial\psi_{H}(r_g)}
      {\partial r_g}
\right)
\delta r_g.
\ee
Since the Hawking temperature \eqref{hawking_temperature} is given by $T_{\textsf{H}({\rm H})}=\frac{\sqrt{h'(r_g)f'(r_g)}}{4\pi}=\frac{\sqrt{h_1(r_g)f_1(r_g)}}{4\pi}$, the differential of the BH entropy is given by 
\be
&&
T_{\textsf{H}({\rm H})}
\delta S_{\rm H}
\nonumber\\
&=&
\int d\Omega
\left(-
\delta
\left(
r^2
\sqrt{\frac{h}{f}}
K^{[tr]}_{(\xi)}
\right)
-
r^2
\sqrt{\frac{h}{f}}
J^{[t}\xi^{r]}
\right)_{r\to r_g}
=
\frac{\sqrt{f_1(r_g) h_1(r_g)}}
{2 G}
\left(
r_g
+
32
\pi G k^{(1)}[\psi_{H} (r_g)]
\frac{\partial\psi_{H}(r_g)}
      {\partial r_g}
\right)
\delta r_g,
\ee
and hence 
\be
\label{deltas_esgb}
\delta S_{\rm H}
=
\frac{2\pi}{G}
\left(
r_g
+
32
\pi G k^{(1)}[\psi_{H} (r_g)]
\frac{\partial\psi_{H}(r_g)}
      {\partial r_g}
\right)
\delta r_g.
\ee
By integrating it, we obtain the BH entropy
\be
\label{Wald_GB}
S_{\rm H}
=
\frac{\pi}{G}
\left(
r_g^2
+
64
\pi G k[\psi_{H} (r_g)]
\right)
=
S_0
\left(
1
+
\frac{64\pi G}{r_g^2}
k[\psi_{H} (r_g)]
\right),
\ee
which agrees with the result by applying standard Wald entropy formula (see e.g., Refs.~\cite{Doneva:2017bvd,Julie:2022huo}). 
Here, we fix a constant in integration such that $\lim_{r_g\rightarrow 0} k[\psi_H(r_g)]=0$. We should note that there is an ambiguity in the definition of the coupling function $k(\phi)$ by adding an arbitrary constant. We can use this freedom to satisfy the above condition.

Thus, the thermodynamic properties of scalarized BHs also remain the same as those argued in the literature~\cite{Doneva:2017bvd,Julie:2022huo}.
We emphasize that although the actions \eqref{sgb} and \eqref{sgb_hor} are equivalent up to the difference in total derivative terms, the dependence of the actions on the spacetime curvature appears to be different.
Nevertheless, the results here indicate that even though the higher-derivative interactions of the scalar field and the nonminimal derivative couplings to the spacetime curvature are present in a description of the theory, by following the original approach of Iyer and Wald and computing the Noether charge potential associated with the diffeomorphism invariance, we could reproduce the results independent of the apparent difference in the action by the total derivative terms. 
We would like to emphasize that not only in the case of the EsGB theories but also in the case of other classes of the Horndeski theories, we should have to obtain the same value of the BH entropy from the two different descriptions of the same theory, whose actions differ by the total derivative terms.
For instance, we have explicitly confirmed that the coincident entropy of a BH solution can be obtained in the two different  descriptions of the same class of the Horndeski theories given by $\left(G_4,G_5\right)= \left(\frac{1}{16\pi G}+c' X, 0\right)$ and $\left(G_4,G_5\right)=\left(\frac{1}{16\pi G}, -c' \phi\right)$ with $c'$ being constant, which are equivalent to each other up to the total derivative terms and also equivalent to the scalar-tensor model with the nonminimal derivative coupling to the Einstein tensor $c' G^{\mu\nu}\phi_\mu \phi_\nu$.

Moreover, since the original action \eqref{sgb} does not include the higher-derivative interactions and the nonminimal derivative couplings of the scalar field to the spacetime curvature, this description may be regarded as the `minimal' one.
Thus, in general we expect that the thermodynamic properties obtained by applying the standard Wald entropy formula to a `minimal' description could be obtained from an equivalent nonminimal description of the same theory including the higher-derivative interactions of the scalar field and/or the nonminimal derivative couplings to the spacetime curvature, by applying  the general scheme employed in this work originally developed by Iyer and Wald.

\subsubsection{\rm non-shift-symmetric EsGB theory}

By solving the set of the equations of motion near the horizon $r=r_g$, $\psi_1$ in Eq.~\eqref{phi_exp} can be found as \cite{Kanti:1995vq,Kanti:1997br,Torii:1996yi}
\be
\psi_1
=\frac{1}{
64\pi G r_g k^{(1)}\left[\psi_{H}(r_g)\right]}
\left[
-r_g^2
+
\sqrt{
r_g^4
-\frac{
1536
\pi G k^{(1)}\left[\psi_{H}(r_g)\right]^2}
      {\eta}}
\right],
\ee
where we choose the branch which recovers the Schwarzschild solution in the limit of $k^{(1)}\left[\psi_{H}(r_g)\right]\to 0$.
Thus, in order for a nontrivial BH solution to exist, we have to impose
\be
\label{existence_condition}
r_g^4
\geq
\frac{
1536
\pi Gk^{(1)}\left[\psi_{H}(r_g)\right]^2}
      {\eta}.
\ee
Let us consider the limit of the absence of the BH horizon $r_g\to 0$.
Assuming the regularity of $\frac{\partial\psi_{H}(r_g)}{\partial r_g}$ in the limit of $r_g\to 0$, i.e., $\psi_{H}(r_g)$ does not blow up as $r_g\to 0$, in the same limit  the second term in the differential \eqref{deltas_esgb} vanishes faster than the first term. Hence, we obtain the vanishing entropy as the usual area law,  by choosing the integration constant so that $S_{\rm H}\to 0$ in the limit of $r_g\to 0$.

In the large distance regions $r\to \infty$, the general vacuum solution in the EsGB theory \eqref{sgb} can be expanded as 
\be
\label{exp_h}
h(r)
&=&
1
-\frac{2{\cal M}(r_g)}{r}
+\frac{
4\pi G\eta {\cal M}(r_g){\cal Q}(r_g)^2}
         {3r^3}
+
{\cal O}
\left(
\frac{1}{r^4}
\right),
\\
\label{exp_f}
f(r)
&=&
1
-\frac{2{\cal M}(r_g)}{r}
+
\frac{
4\pi \eta G {\cal Q}(r_g)^2}
       {r^2}
+\frac{
4\pi G\eta {\cal M}(r_g){\cal Q}(r_g)^2}
         {r^3}
+
{\cal O}
\left(
\frac{1}{r^4}
\right),
\\
\label{exp_psi}
\psi(r)
&=&
\psi_\infty(r_g)
+\frac{{\cal Q}(r_g)}{r}
+
\frac{{\cal M} (r_g){\cal Q}(r_g)}
       {r^2}
-
\frac{{\cal Q}(r_g)\left[-4{\cal M}(r_g)^2+2 \pi G \eta {\cal Q} (r_g)^2\right]}
         {3r^3}
+
{\cal O}
\left(
\frac{1}{r^4}
\right),
\ee
where we assume that the asymptotic amplitude $\psi_\infty(r_g)$, the Arnowitt-Deser-Misner (ADM) mass ${\cal M} (r_g)$, and the scalar charge ${\cal Q}(r_g)$ are the pure functions of the horizon radius $r_g$.
From Eq.~\eqref{ktr_gb}, we obtain the differential of the energy
\be
\label{delta}
\delta M
=
\int d\Omega
\left(-
\delta
\left(
r^2
\sqrt{\frac{h}{f}}
K^{[tr]}_{(\xi)}
\right)
-
r^2
\sqrt{\frac{h}{f}}
J^{[t}\xi^{r]}
\right)_{r\to \infty}
=
\frac{1}{G}
\left(
{\cal M}'(r_g)
+
4\pi G\eta 
{\cal Q} (r_g)
\psi_\infty'(r_g)
\right)
\delta r_g.
\ee
In addition to a non-trivial hairy BH solution, we find a trivial Schwarzschild BH solution with $\phi=\phi_0$ (constant) when the coupling function $k(\phi)$ in the EsGB theory allows  the existence of $\phi_0$ such that $k^{(1)}(\phi_0)=0$. On the other hand, if $k^{(1)}(\phi)\neq 0$ for any values of the scalar field $\phi$, a trivial Schwarzschild spacetime is no longer solution in EsGB theory. We find only a non-trivial hairy BH solution.

\subsubsection{\rm shift-symmetric EsGB theory}

In the shift-symmetric EsGB theories $k(\phi)=\alpha\phi$, where $\alpha$ is the constant, Eq.~\eqref{Wald_GB} reduces to 
\be
\label{shift_symmetric_entropy}
 S_{\rm H}
=
\frac{\pi}{G}
\left(
r_g^2
+
64
\pi G
\alpha \psi_{H} 
\right)
=
S_0
\left(
1
+
\frac{
64
\pi\alpha G}
    {r_g^2}
\psi_{H} 
\right).
\ee
Although even in the shift-symmetric theories the general BH solution could be expanded in the vicinity of the event horizon $r=r_g$ as Eqs.~\eqref{h_exp}-\eqref{phi_exp}, $\psi_{H}$ does not have any physical meaning and hence is not a solution of $r_g$. Thus, we may set the second term in Eq.~\eqref{shift_symmetric_entropy} to zero, by requiring that $S_{\rm H}\to 0$ in the limit of $r_g\to 0$. We then recover  the area law $S_{\rm H}=S_0$ given by Eq.~\eqref{area_law}.
We note that in the shift-symmetric 4D scalar-tensor Einstein-GB theories, recently it was argued that the BH entropy is also given by the area law \cite{Kubiznak:2023xbg}.

In the large distance regions $r\to \infty$, in the expansion \eqref{exp_h}-\eqref{exp_psi}, $\psi_\infty$ also has no physical dependence on $r_g$ in the shift-symmetric theories, and hence Eq.~\eqref{delta} reduces to $\delta M =\frac{{\cal M}'(r_g)}{G}\delta r_g$.
By integrating this, $M=\frac{{\cal M}(r_g)}{G}$, namely the thermodynamic energy coincides with the ADM mass.

\subsection{The irrational coupling model}
\label{sec3d}

Finally, we consider the irrational coupling model
\be
G_2
&=&
\eta X-
\frac{\Lambda}{8\pi G},
\qquad 
G_4
=
\frac{1}{16\pi G}
+\alpha (-X)^{\frac{1}{2}},
\qquad 
G_3
=
G_5
=
0,
\label{irrational}
\ee
Eq. \eqref{integrand__non_qt} reduces to
\be
\label{cq3}
-
\delta
\left(
r^2
\sqrt{\frac{h}{f}}
K^{[tr]}_{(\xi)}
\right)
-
r^2
\sqrt{\frac{h}{f}}
J^{[t}\xi^{r]}
&=&
r^2
\sqrt{\frac{h}{f}}
\Big\{
-\frac{1}{8\pi G r}
\delta f
+
{\cal J}^r
\delta\psi
\Big\}.
\ee
Requiring that the radial component of the Noether current associated with the shift symmetry vanishes,
\be
\label{vanj}
{\cal J}^r
=
-
\eta
f (r)
\psi'(r)
+
\frac{\sqrt{2f(r)}\alpha}
       {r^2}
=0,
\ee
the term proportional to $\delta \psi$ in Eq. \eqref{cq3} vanishes, and hence Eq.~\eqref{cq3} reduces to $-\delta\left(r^2\sqrt{\frac{h}{f}}K^{[tr]}_{(\xi)}\right)-r^2\sqrt{\frac{h}{f}} J^{[t}\xi^{r]}=-\frac{r}{8\pi G} \delta f$.
As the vacuum solution satisfying Eq. \eqref{vanj}, there exists the exact BH solution \cite{Babichev:2017guv}
\be
\label{bcl_solution}
&&
f(r)=h(r)
=
1
-
\frac{
8
\pi G\alpha^2}{r^2\eta}
-
\frac{\Lambda}{3}r^2
-\frac{1}{r}
\left(
r_g
-
\frac{8\pi G \alpha^2}{\eta r_g}
-\frac{\Lambda r_g^3}{3}
\right),
\qquad
\psi'(r)
=
\frac{\sqrt{2}\alpha}{r^2\eta \sqrt{f(r)}}.
\ee
Following the similar steps, evaluating Eq.~\eqref{var_hamilton} with use of Eq.~\eqref{hamiltonian}, we obtain the first law of BH thermodynamics 
\be
T_{\textsf{H}({\rm H})} \delta S_{\rm H}
=
\delta M_{\rm H}
=
\frac{\delta r_g}{2G}
\left(
1-r_g^2\Lambda
+
\frac{8 \pi G \alpha^2}
       {\eta r_g^2}
\right),
\ee
where the Hawking temperature \eqref{hawking_temperature} of the Horndeski BHs is given by 
\be
T_{\textsf{H}({\rm H})}
=
T_0
\left(
1-r_g^2\Lambda
+
\frac{8\pi G
 \alpha^2}
       {\eta r_g^2}
\right).
\ee
By integrating $\delta S_{\rm H}=2 \pi \delta r_g/G$, we obtain the area law Eq.~\eqref{area_law}, where we set the integration constant so that we have $S_{\rm H}\to 0$ in the limit of the vanishing horizon radius $r_g\to 0$.
The mass of the system is given by 
\be
\label{bcl_mass}
M_{\rm H}
=
M_0
\left(
1
-\frac{8
\pi G\alpha^2}{\eta r_g^2}
-\frac{\Lambda r_g^2}{3}
\right),
\ee
which coincides with the total mass of the BH.
We note that the contribution of the scalar field to the ADM mass, which corresponds to the second term in Eq.~\eqref{bcl_mass}, is always negative, as long as the kinetic term of the scalar field has the correct sign $\eta>0$. This  indicates the onset of the ghost instability, which has been observed in the linear stability analysis of the solution \eqref{bcl_solution} performed
in Ref.~\cite{Minamitsuji:2022mlv}.

Since the Schwarzschild -(A)dS metric with the trivial scalar field is not a solution in the theory \eqref{irrational}, there is no other counterpart to compare thermodynamic quantities.

\section{Black holes with linearly time-dependent scalar field}
\label{sec4}

\subsection{Shift- and reflection-symmetric  theories without cosmological constant}
\label{sec4A}

We first focus on subclass of the shift- and reflection-symmetric Horndeski theories, which is invariant under the transformations $\phi\to \phi+c$ with $c$ being constant and $\phi\to -\phi$, and explicitly given by 
\be
\label{shift_ref}
G_2=G_2(X),
\qquad 
G_4=G_4(X),
\qquad 
G_3=G_5=0.
\ee
We assume the static and spherically symmetric spacetime \eqref{metric} and the linearly time-dependent scalar field \eqref{linear_time_sc}.
In this case, Eq.~\eqref{integrand_qt} reduces to 
\be
\label{integrand_qt3}
-
\delta
\left(
r^2
\sqrt{\frac{h}{f}}
K^{[tr]}_{(\xi)}
\right)
-
r^2
\sqrt{\frac{h}{f}}
J^{[t}\xi^{r]}
&=&
r^2
\sqrt{\frac{h}{f}}
\Big\{
-
\frac{2 \delta f}{r}
G_4
-
\frac{4f \psi'}{r}
\left(
f \delta \psi'
+\psi' \delta f
\right)
G_{4X}
\nonumber\\
&&
+
\frac{2f^2\psi'{}^2}{r h^2}
\left\{
q^2 \delta h
+
h^2
\left(
2f  \psi' \delta \psi'
+\psi'{}^2 \delta f
\right)
\right\}
G_{4XX}
\Big\}.
\ee

We focus on the stealth Schwarzschild solution~\cite{Babichev:2013cya,Kobayashi:2014eva}, given by 
\be
\label{stealtH_sol}
&&
f=h=1-\frac{r_g}{r},
\qquad 
X=\frac{q^2}{2},
\qquad
\psi(r)
=
2q\sqrt{r_g}
\left[
\sqrt{r}
-
\sqrt{r_g}
{\rm arctanh}
\left(
\sqrt{\frac{r_g}{r}}
\right)
\right],
\ee
which exists under the conditions
\be
\label{g2g4cond}
G_2
\left(\frac{q^2}{2}\right)
=
G_{2X}
\left(\frac{q^2}{2}\right)
=0.
\ee
Evaluating Eq.~\eqref{var_hamilton} with use of Eq.~\eqref{hamiltonian},
we obtain the first law of thermodynamics 
\be
T_{\textsf{H}({\rm H})} 
\delta S_{\rm H}
=
\delta M_{\rm H}
=
8
\pi
\left(
G_4
\left(\frac{q^2}{2}\right)
-q^2 
G_{4X}
\left(\frac{q^2}{2}\right)
\right)
\delta r_g.
\ee
Since the Hawking temperature \eqref{hawking_temperature} is given by $T_{\textsf{H}({\rm H})}=T_0=\frac{1}{4\pi r_g}$, by integrating $\delta S_{\rm H}$ with respect to $r_g$, we obtain the BH entropy
\be
\label{s_stealth}
S_{\rm H}
=
16
\pi G S_0\left(
G_4\left(\frac{q^2}{2}\right)
-
q^2
G_{4X}\left(\frac{q^2}{2}\right)
\right)
\ee
where $S_0$ is defined in Eq. \eqref{S0}, and we set the integration constant so that we have the vanishing BH entropy $S_{\rm H}\to 0$ in the limit of the vanishing horizon radius $r_g\to 0$.
On the other hand, by integrating $\delta M_{\rm H}$ with respect to $r_g$, we obtain the mass of the system
\be
\label{m_stealth}
M_{\rm H}
=
16
\pi G M_0
\left(
G_4\left(\frac{q^2}{2}\right)
-
q^2
G_{4X}\left(\frac{q^2}{2}\right)
\right).
\ee
where $M_0$ is defined  in Eq. \eqref{M0}, and we set the integration constant so that we have the vanishing mass $M_{\rm H}\to 0$ in the limit of the vanishing horizon radius $r_g\to 0$.

For a more explicit comparison, we consider the specific model \cite{Babichev:2013cya},
\be
G_2(X)=0,
\qquad 
G_4(X)=
\frac{1}{16\pi G}
+\beta X,
\label{stealth_class}
\ee
which trivially satisfies the conditions \eqref{g2g4cond}.
The BH entropy and total mass of the system of the stealth Schwarzschild solutions, Eqs.~\eqref{s_stealth} and \eqref{m_stealth} respectively, reduce to
\be
\label{stealth_entropy_mass}
S_{\rm H}
=
S_0
\left(
1
-
8
\pi G q^2\beta
\right),
\qquad
M_{\rm H}
=
M_0
\left(
1
- 
8
\pi G q^2\beta  
\right).
\label{stealth BH}
\ee
In the same theory \eqref{stealth_class},
there is also the GR Schwarzschild solution with the trivial scalar field, with the BH entropy and mass
\be
\label{gr_sch_entropy_mass}
S_{\rm GR}=
S_0,
\qquad 
M_{\rm GR}
=
M_0,
\ee
respectively.
We will discuss the thermodynamic properties of the stealth Schwarzschild solutions in Sec. \ref{sec5a}.

\subsection{Shift- and reflection-symmetric  theories with cosmological constant (\texorpdfstring{$\Lambda\neq 0$}{TEXT})
}
\label{sec4B}

We focus on the specific shift- and reflection-symmetric subclass of the Horndeski theories \eqref{shift_ref} such that
\be
G_2(X)=\eta X
-\frac{\Lambda}{8\pi G},
\qquad 
G_4(X)=
\frac{1}{16\pi G}
 +\beta X,
\label{sds_class}
\ee
where $\Lambda$ is the cosmological constant and $\beta$ is the coupling constant.
We assume that the static and spherically symmetric spacetime \eqref{metric} and the linearly time-dependent scalar field \eqref{linear_time_sc}.
Then, Eq.~\eqref{integrand_qt} reduces to 
\be
-
\delta
\left(
r^2
\sqrt{\frac{h}{f}}
K^{[tr]}_{(\xi)}
\right)
-
r^2
\sqrt{\frac{h}{f}}
J^{[t}\xi^{r]}
&=&
r^2
\sqrt{\frac{h}{f}}
\frac{1}{r^2h}
\Big\{
-4r\beta f^2h \psi'\delta\psi'
-r
\left[
q^2\beta
+
h \left(
\frac{1}{8\pi G}
+
3\beta f\psi'^2\right)
\right]
\delta f
\Big\}.
\ee
There exist the Schwarzschild-(A)dS solutions
\be
f(r)=
h(r)
=
1-
\frac{\bar \Lambda}{3}
r^2
-\frac{r_g}{r} \left(
1
-
\frac{\bar \Lambda}{3}
r_g^2\right)
,
\qquad 
\psi'(r)
=
q
\frac{\sqrt{1-h(r)}}{\sqrt{f(r)h(r)}},
\ee
with the conditions
\be
\label{def_varlambda}
{\bar \Lambda}:=-\frac{\eta}{2\beta},
\qquad
q=\sqrt{\frac{\eta+2\beta \Lambda}
       {16\pi G \beta \eta}}.
\ee
Thus,
for $\beta>0$ ($\beta<0$),
we obtain the Schwarzschild-AdS (dS) solutions.

Assuming that $\eta>0$ and $G>0$,
for non-negativity inside the square root of $q$, we require 
\be
\Lambda\geq 
{\bar \Lambda}.
\label{two_lambdas}
\ee
Evaluating Eq.~\eqref{var_hamilton} with use of Eq.~\eqref{hamiltonian},
we obtain the first law of BH thermodynamics 
\be
T_{\textsf{H}({\rm H})}
\delta S_{\rm H}
=
\delta M_{\rm H}
=
-
\frac{(2\beta+r_g^2\eta)(-\eta+2\beta\Lambda)}
       {8G\beta\eta}
\delta r_g,
\ee
the Hawking temperature \eqref{hawking_temperature} is given by 
\be
T_{\textsf{H}({\rm H})}
=
\frac{2\beta+r_g^2\eta}
    {8\pi r_g\beta}
=T_0
\left(1+
\frac{\eta}{2\beta}
r_g^2\right).
\ee
Thus, the BH entropy 
for the Schwarzschild-(A)dS solutions
in the Horndeski theory is given by 
\be
S_{\rm H}
=
\frac{\pi r_g^2}{2\eta G} 
\left(\eta-2\beta \Lambda\right)
=
\frac{S_0}{2}
\left(1
-
\frac{2\beta}{\eta}
\Lambda\right),
\ee
where we set the integration constant
so that we have $S_{\rm H}\to 0$ in the limit of the vanishing horizon radius $r_g\to 0$. 
In this theory \eqref{sds_class}, there is also the GR Schwarzschild--(A)dS solution, 
\be
\label{gr_bh}
f_{\rm GR}(r)=
h_{\rm GR}(r)
=
1-
\frac{\Lambda}{3}
r^2
-\frac{r_g}{r}\left(1-
\frac{\Lambda}{3}
r_g^2\right),
\qquad 
\psi'(r)
=
0,
\ee
irrespective of $\beta$ and $\eta$, the BH entropy is given by $S_{\rm GR}=S_0$.
In the limit of $\Lambda={\bar \Lambda}$,where $q=0$ and the scalar field is trivial, we recover the Schwarzschild-(A)dS solutions in GR and obtain the area law \eqref{area_law}.

On the other hand, the mass of the system is given by 
\be
M_{\rm H}=
\frac{r_g\left(
\eta-2\beta\Lambda
\right)
\left(
r_g^2\eta
+6\beta
\right)}
{24G\beta\eta}
=
\frac{M_0}{2}
\left(1-
\frac{2\beta}{\eta}
\Lambda\right)\left(1
+
\frac{\eta}{6\beta}
r_g^2\right),
\ee
where we set the integration constant so that we have $M_{\rm H}\to 0$ in the limit of the vanishing horizon radius $r_g\to 0$, which disagrees with the total mass of the stealth BH given by 
\be
M_{\rm BH}
=
\frac{r_g}{12\beta G}
\left(
r_g^2\eta
+6\beta
\right)
=
M_0
\left(
1-\frac{{\bar \Lambda}}{3} r_g^2
\right),
\ee
except for $\Lambda={\bar \Lambda}$.
For the GR Schwarzschild-(A)dS BHs \eqref{gr_bh}, we obtain $M_{{\rm GR}}=M_0\left(1-\frac{\Lambda}{3} r_g^2\right)$.
We will discuss the thermodynamic properties of the Schwarzschild-(A)dS solutions in Sec. \ref{sec5b}.

\subsection{Shift-symmetric theories with
the coincident speeds of GWs with the speed of light}
\label{sec4C}

Finally, we focus on the subclass of the shift-symmetric Horndeski theories satisfying the requirement that the propagation speed of GWs is equal to the speed of light, i.e., $c_{\rm gw}=c$, whose Lagrangian density is given by 
\be
{\cal L}
=
\frac{1}{16\pi G}
 R
+
G_2(X)
-
G_3(X)
\Box\phi.
\ee
We assume the static and spherically symmetric spacetime \eqref{metric} and the linearly time-dependent scalar field \eqref{linear_time_sc}.
In this case, Eq.~\eqref{integrand_qt} reduces to 
\be
\label{integrand_ct1}
-
\delta
\left(
r^2
\sqrt{\frac{h}{f}}
K^{[tr]}_{(\xi)}
\right)
-
r^2
\sqrt{\frac{h}{f}}
J^{[t}\xi^{r]}
&=&
r^2
\sqrt{\frac{h}{f}}
\Big\{
\Big(
-
\frac{1}{8 \pi G r}
-
\frac{f}{2}
G_{3X}
\psi'^3
\Big)
\delta f
-\frac{q^2 f\psi'}
          {2h^2}
G_{3X}
\delta h
-
f^2
G_{3X}\psi'^2\delta \psi'
\Big\}.
\ee
We focus on the stealth Schwarzschild solution \eqref{stealtH_sol} which exists under the conditions
\be
\label{stealtH_conditions2}
G_2
\left(\frac{q^2}{2}\right)
=G_{2X}
\left(\frac{q^2}{2}\right)
=G_{3X}
\left(\frac{q^2}{2}\right)
=0,
\ee
where Eq.~\eqref{integrand_ct1} further reduces to 
\be
-
\delta
\left(
r^2
\sqrt{\frac{h}{f}}
K^{[tr]}_{(\xi)}
\right)
-
r^2
\sqrt{\frac{h}{f}}
J^{[t}\xi^{r]}
=
r^2
\sqrt{\frac{h}{f}}
\frac{1}{8\pi G r}
\delta f.
\ee
Evaluating Eq.~\eqref{var_hamilton} with use of Eq.~\eqref{hamiltonian}, we obtain the first law of thermodynamics $T_{\textsf{H}({\rm H})} \delta S_{\rm H}=\delta M_{\rm H}=\frac{dr_g}{2G}$, where $T_{\textsf{H}({\rm H})}=T_0$ from Eq. \eqref{hawking_temperature}.
By integrating them, the mass and entropy are, respectively, given by 
\be
M_{\rm H}
=M_0,
\qquad
S_{\rm H}
=S_0,
\ee
where we set the integration constant so that we have $S_{\rm H}\to 0$ and $M_{\rm H}\to 0$ in the limit of  $r_g\to 0$.
Hence we conclude that both BH solutions are equally stable from the viewpoint of BH thermodynamics.

\section{Thermodynamical Instability of Black holes with linearly Time-dependent Scalar Field}
\label{sec5}

As for application of the results presented in Sec.\ref{sec4}, we discuss thermodynamical instability by use of the BH entropy. When some gravitational theory contains two (or more) BH solutions, the comparison of the BH entropy will tell us which BH solution is thermodynamically favored. In GR, the uniqueness of the Kerr(-Newman) BH solution is not held when we include non-Abelian field and/or other fields.  In this case, there exist hairy BH solutions such as colored BHs~\cite{Bartnik:1988am,Bizon:1990sr,Volkov:1990sva,1990JMP....31..928K}. We can then study their stability by a perturbation analysis whose result is consistent with the simple argument by thermodynamical analysis, that is, if the entropy of the first BH solution is smaller than that of the second BH solution, at least the first black hole is thermodynamically unstable~\cite{BIZON1991173, STRAUMANN199033, Maeda:1993ap,Torii:1994nm,Tachizawa:1994wn}.

In the present Horndeski theories, we may discuss thermodynamical instability when there exist two or more  BH solutions. 
As we discussed in Sec.\ref{sec4}, there are BH solutions with linearly time-dependent scalar field. In this section, we discuss thermodynamical stability of those BHs.
Especially, for the BH solutions discussed in Sec. \ref{sec4A} and Sec. \ref{sec4B}, it is argued that perturbations around them are infinitely strongly coupled, and the linear perturbation theory could not be trusted at the arbitrary low energy scales \cite{deRham:2019gha,Motohashi:2019ymr}.
Thus, the stability of the stealth solution is unclear at the level of the linearized analysis. 
However, through the analysis presented in this subsection, we will mention thermodynamical instabilities of these solutions.

\subsection{Shift- and reflection-symmetric  theories without cosmological constant (\texorpdfstring{$\Lambda = 0$)}{TEXT}}
\label{sec5a}

In this subsection, we consider the Horndeski theory given by Eq.~\eqref{stealth_class}.
When we assume a linearly time-dependent scalar field, there exist two Schwarzschild solutions; one is GR Schwarzschild BH with the mass and entropy given by Eq.~\eqref{gr_sch_entropy_mass} and the other is the stealth Schwarzschild BH (the Horndeski Schwarzschild BH) with mass and entropy given by \eqref{stealth_entropy_mass}.
When we compare these two  entropy ($S_{\rm GR}\,,S_{\rm H}$) at the same mass value $M_{\rm GR}=M_{\rm H}$, we can easily find 
\begin{eqnarray}
S_{\rm H}
=
\frac{S_{\rm GR}}
    {1-8\pi G q^2\beta}.
\end{eqnarray}
Hence we obtain the following results:
\begin{eqnarray}
\left\{
\begin{array}{lcl}
S_{\rm H}
&
>S_{\rm GR}
\qquad 
{\rm when} 
\qquad 
\beta>0,
\\
S_{\rm H}
&
 <S_{\rm GR}
\qquad 
{\rm when}
\qquad 
 \beta<0.
\end{array}
\right.
\end{eqnarray}
As a result, we conclude that the Horndeski Schwarzschild BH is thermodynamically stable than the GR Schwarzschild BH 
when $\beta>0$,
while the result turns to be opposite if $\beta<0$.

\subsection{Shift- and reflection-symmetric  theories with cosmological constant
\texorpdfstring{($\Lambda\neq 0$)}{TEXT}}
\label{sec5b}

Here we discuss the Horndeski theory given by Eq.~\eqref{sds_class}.
When we assume a linearly time-dependent scalar field, there exist the two Schwarzschild-(A)dS solutions; one is Schwarzschild-(A)dS solution with the cosmological constant $\Lambda$, and the other is that with the effective cosmological constant $\bar \Lambda=-\eta/2\beta$ given by Eq.~\eqref{def_varlambda}. We have two different Schwarzschild-(A)dS solutions as discussed before, and summarize the thermodynamical variables for two BH solutions as follows:
\begin{itemize}
\item 
Schwarzschild-(A)dS solution with  $\Lambda$ (GR Schwarzschild-(A)dS BH)
\begin{equation}
\label{mass_temperature_schads}
{\rm mass}: M_{\rm GR}=
M_0
\left(1-
\frac{\Lambda r_g^2}{3}
\right)\,,~~ 
{\rm entropy}: S_{\rm GR}=
S_0
\,,~~
{\rm temperature}: T_{\textsf{H}{\rm (GR)}}=
\left(1-\Lambda r_g^2\right)
T_{0}
\,,~~
\end{equation}
where $M_0$ and $S_0$ are defined in Eqs. \eqref{M0} and \eqref{S0}, and $T_{0}:=\frac{1}{4\pi r_g}$ represents the Hawking temperature in the Schwarzschild background with the horizon radius $r_g$.
\item 
Schwarzschild-(A)dS solution with  $\bar \Lambda$ (Horndeski Schwarzschild-(A)dS BH)
\begin{equation}
{\rm mass}: M_{\rm H}=
M_0
\left(1-
\frac{{\bar \Lambda} r_g^2}{3}
\right)
\frac{\Lambda+{\bar\Lambda}} 
      {2\bar\Lambda}\,,~~
{\rm entropy}: S_{\rm H}=
S_0
\frac{\Lambda+\bar\Lambda} 
    {2\bar\Lambda}
\,,~~
{\rm temperature}: 
T_{\textsf{H}({\rm H})}=
\left(1-\bar \Lambda r_g^2\right)
T_0
\,.~~
\end{equation}
\end{itemize}
Since $\Lambda\geq\bar\Lambda$, we can classify the solutions into three cases: (1) $\Lambda\geq \bar\Lambda>0$~($\beta<0$),  (2) $\Lambda>0\,, \bar\Lambda<0$~($\beta>0$), (3) $0>\Lambda\geq \bar\Lambda$~($\beta>0$). In the case (1), two BH solutions are Schwarzschild-dS solutions, while in the case (3), we find two Schwarzschild-AdS solutions. We shall discuss their theomodynamical instabilities in below. 
For the case (2), since one is Schwarzschild-dS solution and the other is Schwarzschild-AdS solution, the boundary conditions are completely different. We may not expect any phase transition between them.
We introduce the curvature radii $\ell$ and $\bar \ell$, which are defined by $\ell=\sqrt{3/\epsilon_{\Lambda}\Lambda}$ and $\bar \ell =\sqrt{3/\epsilon_{\bar \Lambda}\bar \Lambda}$, where $\epsilon_{\Lambda}$ and $\epsilon_{\bar \Lambda}$ are the signs of $\Lambda$ and $\bar \Lambda$, respectively.

\subsubsection{\texorpdfstring{$\Lambda\ge \bar\Lambda>0$}{TEXT}}
\begin{figure}[ht]
\begin{center}
\includegraphics[width=6cm]{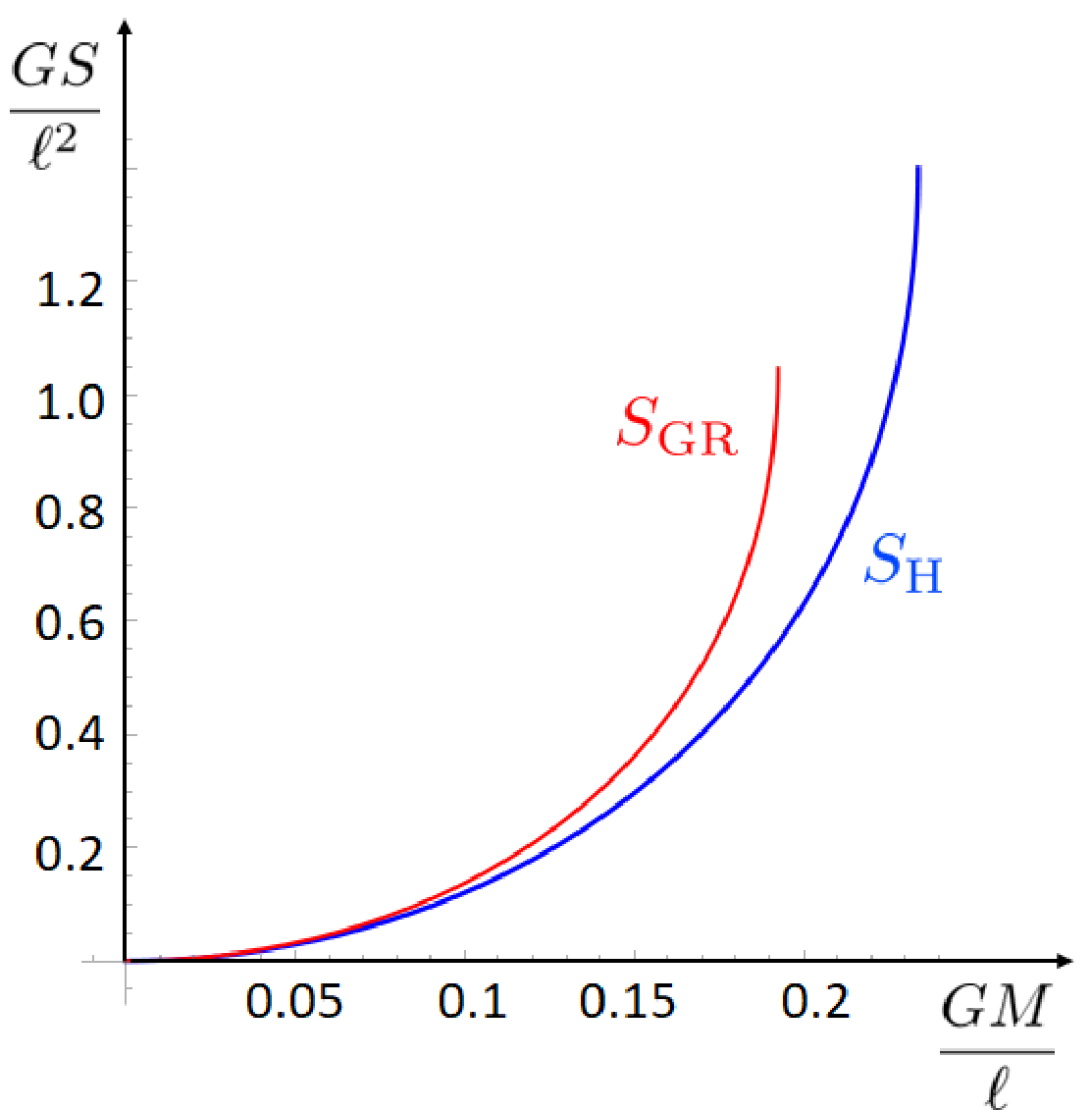}
\caption{The entropy of the 
GR Schwarzschild-dS BH (the red curve) and that of the 
Horndeski Schwarzschild-dS BH
(the blue curve) in terms of the mass.}
\label{SdS}
\end{center}
\end{figure}

In this case, $\ell\leq \bar\ell$, and the thermodynamical valuables are given by 
\begin{eqnarray}
&&M_{\rm GR}=
M_0
\left(1-
\frac{r_g^2}{\ell^2}
\right)\,,~~ 
S_{\rm GR}=
S_0 \,,~~
T_{\textsf{H}{\rm (GR)}}=
\left(1-
\frac{3r_g^2}{\ell^2}
\right)
T_{0}
\,,~~
\label{SdS_GR}
\\
&&M_{\rm H}=
M_0
\left(1-
\frac{r_g^2}{\bar \ell^2}
\right)
\frac{\ell^2+\bar\ell^2}{2\ell^2}
\,,~~
S_{\rm H}=
S_0
\frac{\ell^2+\bar\ell^2}{2\ell^2}
\,,~~
T_{\textsf{H}({\rm H})}=
\left(1-
\frac{3r_g^2}{{\bar \ell}^2}
\right)
T_{0}\,.
\label{SdS_H}
\end{eqnarray}
In order to discuss thermodynamical stability, we plot the mass-entropy diagram, which is given in Fig. \ref{SdS} for the case of ${\bar \ell}/\ell=1.1$.

For a given mass $M$, the entropy of the GR Schwarzschild-dS BH with $\Lambda$ is always larger than that of the Horndeski Schwarzschild-dS BH  with $\bar \Lambda$.
It means that the Horndeski Schwarzschild-dS BH is thermodynamically unstable than the GR Schwarzschild-dS BH. 
We expect that thermodynamical phase transition from the Horndeski Schwarzschild-dS BH to the GR Schwarzschild-dS BH.
Since there exists a scalar field $\phi$ outside the Horndeski Schwarzschild-dS BH, the scalar field propagates away to infinity when the transition occurs.
If the entropy is conserved, the mass energy decreases by the emission of a scalar field.
In general, we expect that the entropy increases as well as the mass energy decreases 
and the Horndeski Schwarzschild-dS BH transits to the GR Schwarzschild-dS BH in the left-up direction in the diagram.

\subsubsection{\texorpdfstring{${\bar\Lambda}\le \Lambda<0$}{TEXT}}

In this case, both BHs are described by the Schwarzschild-AdS solutions with $\ell\geq \bar\ell$, and 
the thermodynamical valuables are given by 
\begin{eqnarray}
&&M_{\rm GR}=
M_0
\left(1+
\frac{r_g^2}{\ell^2}
\right)\,,~~ 
S_{\rm GR}=
S_0\,,~~
T_{\textsf{H}{\rm (GR)}}=
\left(1+
\frac{3r_g^2}{\ell^2}
\right)T_0\,,~~
\label{SAdS_GR}
\\
&&M_{\rm H}=
M_0
\left(
1+
\frac{r_g^2}{{\bar \ell}^2}
\right)
\frac{\ell^2+\bar\ell^2}{2\ell^2}
\,,~~
S_{\rm H}=
S_0
\frac{\ell^2+\bar\ell^2}{2\ell^2}
\,,~~
T_{\textsf{H}({\rm H})}= 
\left(1+
\frac{3r_g^2}{{\bar \ell}^2}
\right)
T_0\,.
\label{SAdS_H}
\end{eqnarray}

\begin{figure}[ht]
\begin{center}
\includegraphics[width=5cm]{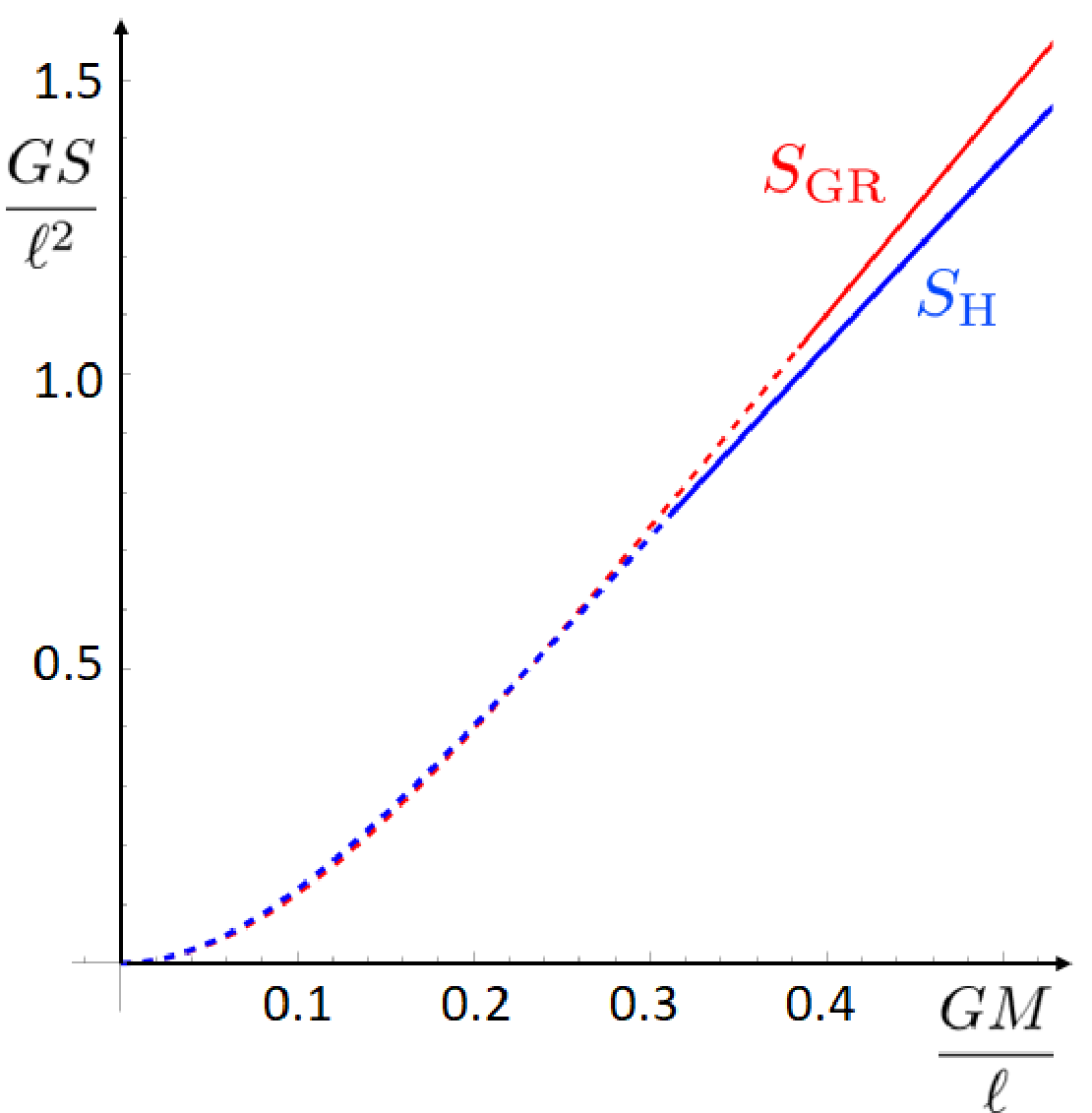}~~~~
\includegraphics[width=5cm]{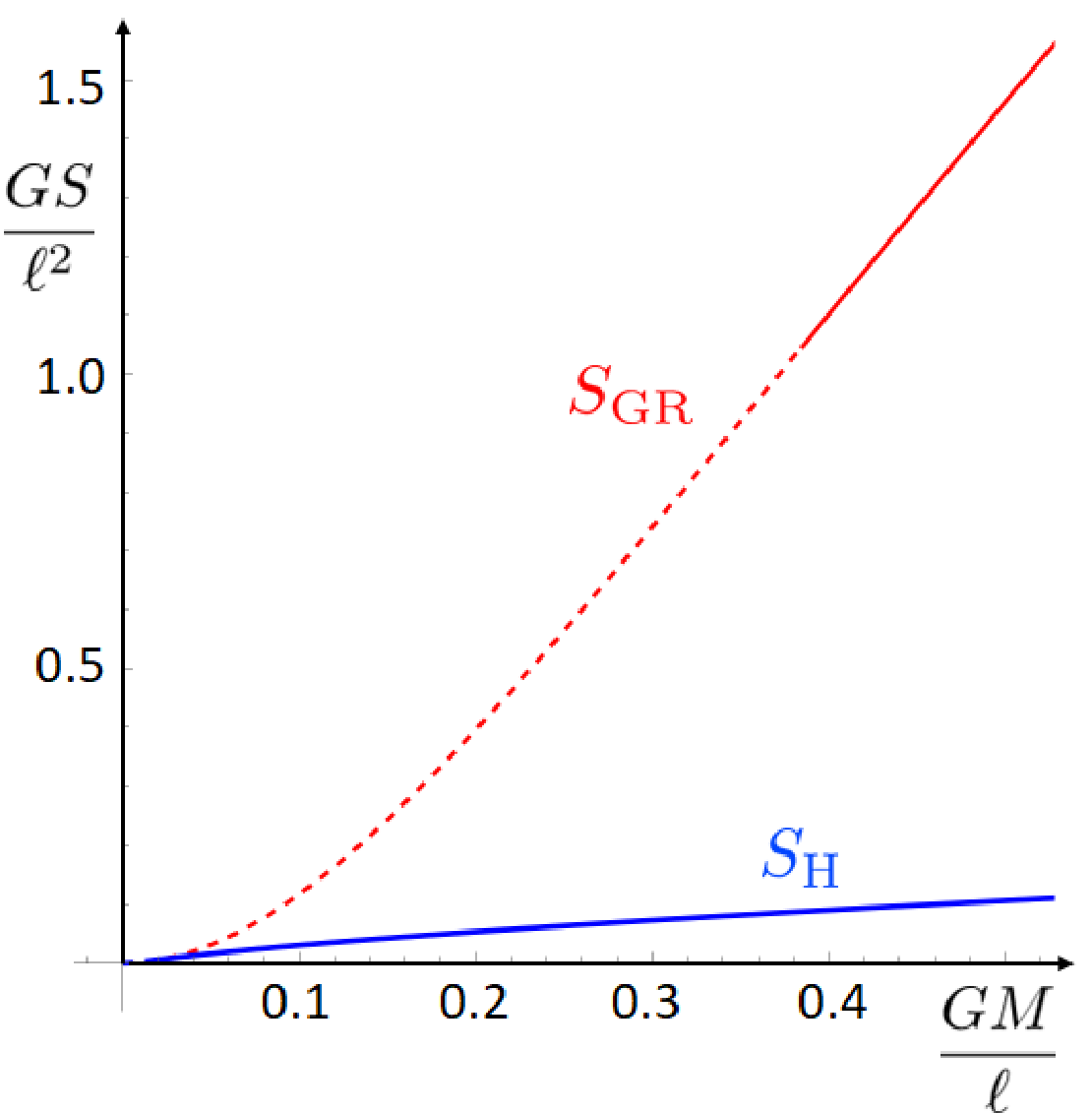}~~
\includegraphics[width=5cm]{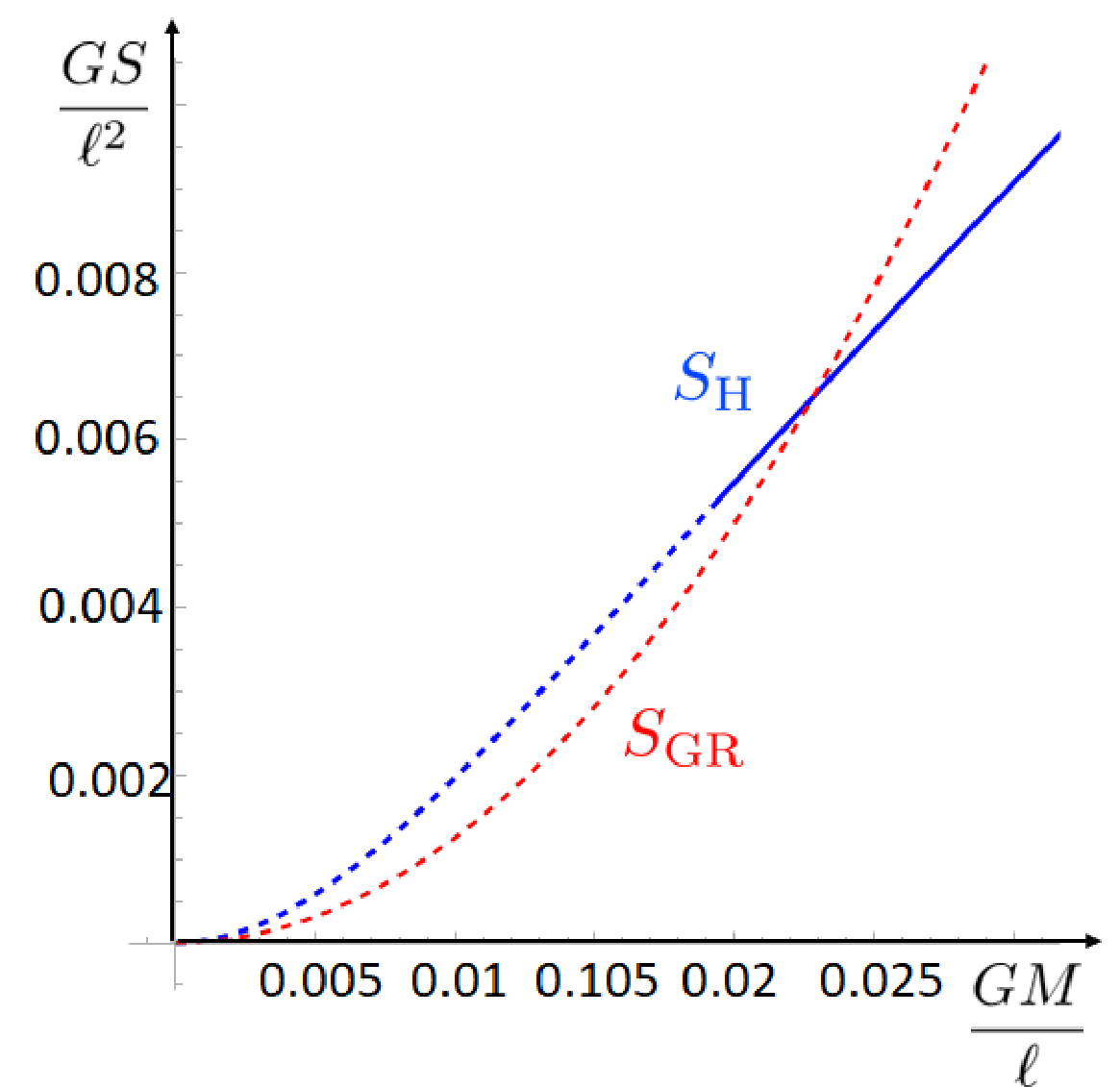}
\\
(a)\hskip 5.5cm (b) \hskip 5cm (c)
\end{center}
\caption{The entropy of the GR Schwarzschild-AdS BH (the red curve) and that of the Horndeski Schwarzschild-AdS BH (the blue curve) in terms of the mass ((a) is the case of $\bar \ell/\ell=0.9$, while (b) is for $\bar \ell/\ell=0.1$. (c) is the enlarged version of (b)). 
The dotted curves are thermal AdS phases via Hawking-Page transition, while the solid curves denote "large" Schwarzschild-AdS BH phases. }
\label{SAdS}
\end{figure}

We plot the mass-entropy diagram, which is given in Fig. \ref{SAdS} for the case of $\bar \ell/\ell=0.9$ (a) and $0.1$ (b), (c).

In this case, two curves $S_{\rm GR}(M)$ and $S_{\rm H}(M)$ intersect at some critical mass $M_{\rm GR\mathchar`-H}$, beyond which $S_{\rm GR}>S_{\rm H}$. In an asymptotically AdS spacetime, there exists another critical mass $M_{\textsf{HP}}$, below which Schwarzschild-AdS BH evaporates to thermal radiation in AdS space via the Hawking-Page transition~\cite{Hawking:1982dh}. 
In the present case, since there are two Schwarzschild-AdS BH solutions, we find two critical masses, $M_{\textsf{HP} ({\rm GR})}$ and $M_{\textsf{HP} ({\rm H})}$ corresponding to the GR Schwarzschild-AdS BH and the Horndeski Schwarzschild-AdS BH, respectively. We find that $M_{\textsf{HP} ({\rm GR})}>M_{\textsf{HP} ({\rm H})}$. 
In the limit of $\bar \ell\rightarrow \ell$, the critical horizon radius $r_{g{\rm GR\mathchar`-H}}$ becomes $\ell/\sqrt{5}$, which is smaller than the HP transition radius $r_{g\textsf{HP} ({\rm GR})}=\ell/\sqrt{3}$. We then find $M_{\rm GR\mathchar`-H}<M_{\textsf{HP} ({\rm GR})}$.
As a result, we can classify into two cases; (1) $M_{\textsf{HP} ({\rm GR})}>M_{\textsf{HP} ({\rm H})}>M_{\rm GR\mathchar`-H}$, and  (2) $M_{\textsf{HP} ({\rm GR})}>M_{\rm GR\mathchar`-H}>M_{\textsf{HP} ({\rm H})}$.
If $\mathfrak{r}_{\rm cr}<\bar \ell/\ell <1$, we find the case (1), while when $0<\bar \ell/\ell <\mathfrak{r}_{\rm cr}$, we obtain the case (2). 
The critical value $\mathfrak{r}_{\rm cr}$ is given by the root of the equation $\mathfrak{r}_{\rm cr}^6 + 3 \mathfrak{r}_{\rm cr}^4+ 16 \mathfrak{r}_{\rm cr}^2-4  = 0$, i.e. $\mathfrak{r}_{\rm cr}\approx 0.48835$.
We then find the following various evolution scenarios depending the coupling constants:
\begin{itemize}
\item Case (1) $\mathfrak{r}_{\rm cr}<\bar \ell/\ell <1$
\\
Below $M_{\rm GR\mathchar`-H}$, we find only thermal radiation in AdS space, in which the effective cosmological constant is fixed by $\bar \Lambda$, while in the range of $M_{\rm GR\mathchar`-H}<M<M_{\textsf{HP} ({\rm H})}$, it is also thermal radiation in AdS space but with the cosmological constant $\Lambda$. In the range of $M_{\textsf{HP} ({\rm GR})}>M>M_{\textsf{HP} ({\rm H})}$, the Horndeski Schwarzschild-AdS BH will evaporate via the Hawking-Page transition, finding  thermal radiation in AdS space but with the cosmological constant $\Lambda$. When $M>M_{\textsf{HP} ({\rm GR})}$, the Horndeski Schwarzschild-AdS BH will evolve into the GR Schwarzschild-AdS BH via thermal phase transition.
\item Case (2)  $0<\bar \ell/\ell <\mathfrak{r}_{\rm cr}$
\\
Below $M_{\textsf{HP} ({\rm H})}$ we find only thermal radiation in AdS space, in which the effective cosmological constant is fixed by $\bar \Lambda$ just as Case (1). 
In the range of $M_{\textsf{HP} ({\rm H})}<M<M_{\rm GR\mathchar`-H}$, we find the transition from thermal radiation in AdS space with $\Lambda$ into the stable Horndeski Schwarzschild-AdS BH. 
In the range of $M_{\textsf{HP} ({\rm GR})}>M>M_{\textsf{HP} ({\rm H})}$,  the Horndeski Schwarzschild-AdS BH will evaporate into thermal radiation in AdS space with $\Lambda$.
When $M>M_{\textsf{HP} ({\rm GR})}$, the Horndeski Schwarzschild-AdS BH will evolve into the GR Schwarzschild-AdS BH via thermal phase transition just as Case (1).
\end{itemize}

\subsection{Relation between the cases of \texorpdfstring{$\Lambda=0$ (Sec. \ref{sec5a})  and  of $\Lambda\neq 0$ (Sec.\ref{sec5b})}
{TEXT}}
\label{sec5c}

In order to discuss the relation between Schwarzschild BH and Schwarzschild (A)dS BH discussed in the two previous  subsections, we rewrite the mass and entropy by use of $\beta$ and $q$.
From Eq. (\ref{def_varlambda}), we find
\be
\frac{\bar \ell^2+\ell^2}{2\ell^2}
=
\frac{\bar \Lambda+\Lambda}{2{\bar \Lambda}}
=1-
8 \pi G q^2\beta
\ee
Using this relation and Eqs. \eqref{SdS_GR},\eqref{SdS_H}, \eqref{SAdS_GR}, and \eqref{SAdS_H}, we obtain the relation between the masses $M_{\rm GR}$ and $M_{\rm H}$ and that of the entropies $S_{\rm GR}$ and $S_{\rm H}$ as
\begin{eqnarray}
&&M_{\rm GR}=
M_0
\left(1\mp 
\frac{r_g^2}{\ell^2}
\right)\,,~~ 
S_{\rm GR}=
S_0
\\
&&M_{\rm H}=
M_0
\left(1\mp 
\frac{r_g^2}{{\bar \ell}^2}
\right)(1-8\pi G q^2\beta)\,,~~
S_{\rm H}=
S_0
\left(1-8\pi G q^2\beta\right)\,,
\end{eqnarray}
where $\mp$ correspond to the Schwarzschild-dS BH and Schwarzschild-AdS BH, respectively.
When we take the limit of $\Lambda\,,\bar\Lambda \rightarrow 0~(\ell\,,\bar \ell \rightarrow \infty)$, we find the same relation (\ref{stealth BH}) of the masses and entropies of the stealth Schwarzschild BH and GR Schwarzschild BH. 

As discussed in Sec. \ref{sec5b}, for Schwarzschild-AdS BH ($\beta>0$), $S_{\rm H}>S_{\rm GR}$ in the small mass limit, while for the Schwarzschild-dS BH ($\beta<0$) the relation becomes opposite, which is consistent with thermodynamical instability of Schwarzschild BH discussed in Sec. \ref{sec5a}.
We note that Schwarzschild BH does not show the Hawking-Page transition, and then BH with larger entropy becomes stable than the other.

\section{Black hole thermodynamics of charged black holes}
\label{sec6}

BH solutions discussed so far contained only one independent charge, i.e., the mass of the BH, or equivalently the radius of the BH event horizon.
In order to discuss BH thermodynamics with two or more independent charges,
in this section we will focus on the Horndeski theories minimally coupled to the $U(1)$-invariant vector field.
In such theories, BH solutions could contain at least two independent charges, the mass and electric (and/or magnetic) charges.
Here, we will focus on the electrically charged BH solution as an extension of the earlier work \cite{Feng:2015wvb}.

We will extend the general formulation presented in Sec.~\ref{sec2} and derive the Noether change potential associated with the diffeomorphism invariance, including the contribution of the general $U(1)$-invariant vector field. 
We will then apply our formulation to static and spherically symmetric charged BH solutions and obtain the variations of Hamiltonian evaluated at the horizon and in the spatial infinity.
As a concrete example of charged BH solutions, we will consider the extension of the irrational coupling model discussed in Sec.~\ref{sec3d}  minimally coupled to the ordinary Maxwell field.
We show that in this model the differential of the BH entropy is integrable and the first law of the BH thermodynamics is recovered.

We will then consider the general reflection- and shift-symmetric class of the Horndeski theories minimally coupled to the Maxwell field, and clarify the general conditions under which the differential of the BH entropy is integrable in the presence of the two independent charges.

\subsection{The Noether charge potential with the $U(1)$-invariant vector field}

We consider the Horndeski theory minimally coupled to a $U(1)$-invariant vector field
\be
\label{action2}
S
&=&
\int d^4x\sqrt{-g}
 {\cal L}
=
\int d^4x\sqrt{-g}
\left(
\sum_{i=2}^5 {\cal L}_i
+
G_A(\mF)
\right),
\ee
where ${\cal L}_i$ ($i=2,3,4,5$) are given 
given by \eqref{l2}-\eqref{l5}, 
and $G_A(\mF)$ is the general $U(1)$-invariant Lagrangian density for the vector field $A_\mu$
given as the general function of 
\be
\mF
:=-\frac{1}{4}
g^{\alpha\beta} g^{\mu\nu}
F_{\alpha\mu}F_{\beta\nu},
\ee
with $F_{\mu\nu}:=\partial_\mu A_\nu-\partial_\nu A_\mu$ being the electromagnetic field strength.
The variation of the action \eqref{action2} is given by 
\be
\delta S
=
\int d^4x
\sqrt{-g}
\left(
E_{\mu\nu} 
\delta g^{\mu\nu}
+
E_\phi
\delta \phi
+
\nabla_\nu\left(G_{A,\mF} F^{\nu}{}_{\mu}\right)
\delta A^\mu
+
\nabla_\mu J^\mu
\right),
\ee
where the equation of motion of the vector field is given by $\nabla_\nu\left(G_{A,\mF} F^{\nu}{}_{\mu}\right)=0$ with $G_{A,\mF}:= \frac{\partial G_A}{\partial \mF}$, and the boundary current is given by 
\be
J^\mu
&=&
\sum_{i=2}^5 J^\mu_{(i)}
+J_A^\mu,
\ee
which $J^\mu_{(i)}$ ($i=2,3,4,5$) is given by Eq.~\eqref{j2}-\eqref{j5},
and 
\be
J_A^\mu
:= 
-G_{A,\mF} F^{\mu\nu}\delta A_\nu.
\ee
We also define the dual 3-form to $J^\mu$ as Eq. \eqref{dual_3form}

Under the diffeomorphism transformation, $x^\mu\to x^\mu+\xi^\mu (x^\mu)$, the variations of the metric and scalar field are given by Eq. \eqref{diffeo}, and that of the vector field is given by $\delta_\xi A_\mu=\xi^\sigma \nabla_\sigma A_\mu+A_\sigma \nabla_\mu A^\sigma$, respectively.
Using the background equations of motion, under the diffeomorphism transformation, we obtain
\be
\label{jmu_g}
J_{(\xi)}^\mu 
-
\xi^{\mu} 
{\cal L}
=
2
\nabla_\nu K^{[\nu\mu]}_{(\xi)}
=
2
\nabla_\nu
\left(
\sum_{i=2}^5
 K^{[\nu\mu]}_{(i)(\xi)}
+
K^{[\nu\mu]}_{A(\xi)}
\right),
\ee
where each individual contribution from the Horndeski theories is given by Eqs.~ \eqref{k2}-\eqref{k5},
and the Noether charge potential for the vector field is given by
\be
K^{\mu\nu}_{(\xi)A}
&=&
\frac{1}{2}
G_{A,\mF}
F^{\mu\nu}
A_\sigma \xi^\sigma,
\ee
respectively.
We then define the dual 2-form of the Noether charge potential $K_{(\xi)}^{\mu\nu}$ as Eq. \eqref{def_q}, and the 2-form tensor where the first index of ${\Theta}_{\nu\alpha\beta}$ defined in Eq.~\eqref{dual_3form} is contracted by the infinitesimal differmorphism transformation $\xi^\nu$, by Eq. \eqref{itheta}.
We then consider the variation of the dual Noether charge potential with respect to the physical parameters subtracted by Eq.~\eqref{itheta} as Eq. \eqref{deltaQ}.
The integration of Eq.~\eqref{deltaQ} on the boundaries of the Cauchy surface gives rise to the variation of the Hamiltonian \cite{Wald:1993nt,Iyer:1994ys}.

As the background, we consider the static and spherically symmetric solutions whose metric is written by Eq.~\eqref{metric}, where the functions $h(r)$ and $f(r)$ contain the common largest root at $r=r_g>0$ corresponding to the position of the BH event horizon, and $f(r)>0$ and $h(r)>0$ for $r>r_g$.
For the scalar field, we focus on the static ansatz \eqref{static_sc} for simplicity.
For the $U(1)$-invariant vector field, we assume the following ansatz
\be
A_\mu
=
\left(
A_0(r),0,0,0
\right),
\ee
which only gives rise to the electric field $F_{rt}=A_0'(r)$.
We choose the gauge such that the value of $A_0(r)$ vanishes on the horizon $r=r_g$, i.e.,~$A_0(r_g)= 0$.
We assume that $\xi^\mu$ corresponds to the timelike Killing vector field, $\xi^\mu=(1,0,0,0)$.

The variations in terms of the integration constants can be written as Eq. \eqref{variation} and $\delta A_0(r)=\sum_j\frac{\partial A_0(r)}{\partial c_j}\delta c_j$, where $c_j$'s are integration constants of the BH solutions, which include the position of the event horizon $r_g$ and the electric charge $Q$.
With use of Eq.~\eqref{deltaQ}, the variation of the Hamiltonian with respect to the integration constants is given by the contributions from the horizon $r\to r_g$ and infinity $r\to \infty$, $\delta {\cal H}=\delta{\cal H}_\infty-\delta{\cal H}_{H}$, where $\delta{\cal H}_\infty$ and $\delta{\cal H}_{H}$ are given by Eq.~\eqref{hamiltonian}.
The conservation of the Hamiltonian ${\cal H}=0$ yields
\be
\label{conservation}
\delta{\cal H}_\infty=\delta{\cal H}_{H}.
\ee

\subsection{The Einstein-Maxwell theory}

As the simplest example without the dynamical scalar field $\phi=0$, we consider the Einstein-Maxwell theory with the cosmological constant $\Lambda$
\be
G_2
=
-
\frac{\Lambda}{8\pi G}
\qquad
G_4
=
\frac{1}{16\pi G},
\qquad 
G_3
=
G_5
=
0,
\qquad 
G_A
=
\mF,
\label{em_f}
\ee
under which Eq. \eqref{integrand__non_qt} reduces to
\be
\label{cq3_f}
&&
-
\delta
\left(
r^2
\sqrt{\frac{h}{f}}
K^{[tr]}_{(\xi)}
\right)
-
r^2
\sqrt{\frac{h}{f}}
J^{[t}\xi^{r]}
\nonumber
\\
&=&
r^2
\sqrt{\frac{h}{f}}
\Big\{
-
\left[
\frac{1}{8\pi G r}
+
\frac{A_0(r)A_0'(r)}
    {2h(r)}
\right]
\delta f(r)
+
\frac{f(r) A_0(r)A_0'(r)}{2h(r)^2}
\delta h(r)
-
\frac{A_0(r) f(r)}{h(r)}
\delta A_0'(r)
\Big\}.
\ee
In the theory \eqref{em_f},
there exists the Reissner-Nordstr\"{o}m-de Sitter solution
\be
\label{bcl_solution_f}
f(r)
=
h(r)
&=&
1
-\frac{\Lambda}{3}r^2
+
\frac{1}{r}
\left(
\frac{r_g}{3}
\left(
-3+r_g^2\Lambda
\right)
-
\frac{4\pi G Q^2}
        {r_g}
\right)
+
\frac{4\pi G Q^2}
        {r^2},
\quad
A_0(r)
=
Q
\left(
\frac{1}{r}
-
\frac{1}{r_g}
\right),
\ee
which satisfies the gauge condition $A_0(r_g)=0$.

The variation of the Hamiltonian on the horizon $r=r_g$ yields
\be
\label{var_horizon_em}
\delta{\cal H}_H
=
T_{\textsf{H}({\rm H})} \delta S_{\rm H}
=
\frac{\delta r_g}{2G}
\left[
1
-r_g^2\Lambda
-
\frac{4\pi G Q^2}
       {r_g^2}
\right],
\ee
where the Hawking temperature \eqref{hawking_temperature} is given by 
$T_{\textsf{H}({\rm H})}
=
T_0
\left(
1
-r_g^2\Lambda
-
\frac{4\pi GQ^2}
       {r_g^2}
\right)$.
Thus, we obtain the integrable relation $\delta S_{\rm H}=\left(2 \pi r_g/G \right)\delta r_g$, as the proportionality coefficient $\left(2\pi r_g/G\right)$ does not depend on the electric charge $Q$.
As the consequence, we obtain the area law Eq.~\eqref{area_law}, where we set the integration constant so that we have $S_{\rm H}\to 0$ in the limit of the vanishing horizon radius $r_g\to 0$.

The variation of the Hamiltonian at the infinity $r\to \infty$ yields
\be
\label{var_infinity_em}
\delta {\cal H}_\infty
=
\frac{\partial M_H}
        {\partial r_g}
\delta r_g
=
\delta M_H
-
\frac{\partial M_H}{\partial Q}
\delta Q
=
\delta M_H
-
\Phi_H
\delta Q,
\ee
where $\Phi_H:=-
4\pi \left(A_0(r\to \infty)-A_0(r_g)\right)=-
4\pi A_0(r\to \infty)$ describes the difference in the electric potential between the infinity $r\to \infty$
 and the horizon $r=r_g$
\footnote{
If we choose another gauge condition for $A_0(r)$ such that $A_0(r_g)\neq 0$,
the variation of the Hamiltonian at $r=r_g$, $\delta {\cal H}_H$, includes a term proportional to $\delta Q$ as well as the right-hand side of Eq.~\eqref{var_horizon_em}.
However, the same term proportional to $\delta Q$ will also appear in the variation of the Hamiltonian in the limit of  $r\to\infty$, $\delta {\cal H}_\infty$, as well as the terms in Eq.~\eqref{var_infinity_em}. 
Thus, in the conservation of the Hamiltonian Eq.~\eqref{conservation} this gauge-dependent term proportional to $\delta Q$ cancels, leading to the first law of BH thermodynamics as Eq.~\eqref{charged_first_law}, as expected.}, and the mass of the total system is given by 
\be
\label{bcl_mass_f}
M_{\rm H}
=
M_0
\left(
1
+
\frac{4\pi G Q^2}
       {r_g^2}
-\frac{\Lambda r_g^2}{3}
\right),
\ee
which coincides with the total ADM mass of the BH spacetime.
The conservation of the Hamiltonian ${\cal H}=0$, Eq.~\eqref{conservation} yields the first law of the thermodynamics for the electrically charged BH

\be
\label{charged_first_law}
T_{\textsf{H}({\rm H})} \delta S_{\rm H}
=
\delta M_H
-
\Phi_H
\delta Q.
\ee

\subsection{The irrational coupling model with the $U(1)$-invariant vector field}
\label{sec6c}

We then consider the irrational coupling model with the minimally coupled $U(1)$-invariant vector field
\be
G_2
=
\eta X-
\frac{\Lambda}{8\pi G},
\qquad 
G_4
=
\frac{1}{16\pi G}
+\alpha (-X)^{\frac{1}{2}},
\qquad 
G_3
=
G_5
=
0,
\qquad 
G_A
=\mF,
\label{irrational_f}
\ee
under which Eq. \eqref{integrand__non_qt} reduces to
\be
\label{cq3_em}
&&
-
\delta
\left(
r^2
\sqrt{\frac{h}{f}}
K^{[tr]}_{(\xi)}
\right)
-
r^2
\sqrt{\frac{h}{f}}
J^{[t}\xi^{r]}
\nonumber
\\
&=&
r^2
\sqrt{\frac{h}{f}}
\Big\{
-
\left[
\frac{1}{8\pi G r}
+
\frac{A_0(r)A_0'(r)}{2h(r)}
\right]
\delta f(r)
+
\frac{f(r) A_0(r)A_0'(r)}{2h(r)^2}
\delta h(r)
-
\frac{f(r)A_0(r)}
          {
          h(r)}
\delta A_0'(r)
+
{\cal J}^r
\delta\psi(r)
\Big\},
\ee
where the radial component of the Noether current associated with the shift symmetry ${\cal J}^r$ is given by Eq. \eqref{vanj}.
We note that the model \eqref{irrational_f} corresponds to an extension of Eq. \eqref{irrational} with the Maxwell field, discussed in Sec.~\ref{sec3d}.
There exists the exact BH solution with the electric charge $Q$
\be
\label{bcl_solution_f2}
f(r)
=
h(r)
&=&
1
-\frac{\Lambda}{3}r^2
+
\frac{1}{r}
\left(
\frac{r_g}{3}
\left(
-3+r_g^2\Lambda
\right)
-
\frac{4
\pi G}{r_g}
\left(Q^2-\frac{2\alpha^2}{\eta}\right)
\right)
+
\frac{4
\pi G}{ r^2}
\left(
Q^2
-
\frac{2\alpha^2}{\eta}
\right),
\nonumber
\\
\psi'(r)
&=&
\frac{\sqrt{2}\alpha}{r^2\eta \sqrt{f(r)}},
\qquad 
A_0(r)
=
Q
\left(
\frac{1}{r}
-
\frac{1}{r_g}
\right),
\ee
which satisfies $A_0(r_g)=0$.

The variation of the Hamiltonian on the horizon $r=r_g$ yields
\be
\label{var_horizon_irrational_f}
\delta{\cal H}_H
=
T_{\textsf{H}({\rm H})} \delta S_{\rm H}
=
\frac{\delta r_g}{2G}
\left[
1
-r_g^2\Lambda
-
\frac{4\pi G}
       {r_g^2}
\left(
Q^2
-
\frac{2\alpha^2}{\eta}
\right)
\right],
\ee
where the Hawking temperature \eqref{hawking_temperature} is given by
$T_{\textsf{H}({\rm H})}=T_0\left(1-r_g^2\Lambda-\frac{4\pi G}{r_g^2}\left(Q^2-\frac{2\alpha^2}{\eta}\right)\right)$.
As in the case of the Einstein-Maxwell theory, in Eq.~\eqref{var_horizon_irrational_f}, the terms proportional to the variation $\delta Q$ vanish.
We obtain the integrable relation $\delta S_{\rm H}=\left(2 \pi r_g/G\right) \delta r_g$ and the proportionality coefficient $\left(2\pi r_g/G\right)$ does not depend on the electric charge $Q$, and as the consequence, we obtain the area law Eq.~\eqref{area_law} in spite of the existence of the two independent charges.

The variation of the Hamiltonian at the infinity $r\to \infty$ yields Eq. \eqref{var_infinity_em}, 
where $\Phi_H=-4\pi A_0(r\to \infty)$ in our gauge condition describes the difference in the electric potential between the infinity $r\to \infty$ and the horizon $r=r_g$, and the mass of the system is given by 
\be
\label{bcl_mass_f_hor}
M_{\rm H}
=
M_0
\left(
1
+
\frac{4\pi G}
       {r_g^2}
\left(
Q^2
-
\frac{2\alpha^2}{\eta}
\right)
-\frac{\Lambda r_g^2}{3}
\right),
\ee
which coincides with the total ADM mass of the BH spacetime.
The conservation of the Hamiltonian~\eqref{conservation} yields the first law of the thermodynamics for the charged BH \eqref{charged_first_law} as in the Einstein-Maxwell theory.

\subsection{The reflection- and shift-symmetric model with the $U(1)$-invariant vector field}

Finally, we consider the general reflection- and shift-symmetric class of the Horndeski theories with the minimally coupled $U(1)$-invariant vector field
\be
G_2
&=&
g_2(X),
\qquad 
G_4
=
\frac{1}{16\pi G}
+
g_4(X)
\qquad 
G_3
=
G_5
=
0,
\qquad 
G_A
=\mF,
\label{irrational_f2}
\ee
where $g_2(X)$ and $g_4(X)$ are general functions of the kinetic term $X$,
under which Eq. \eqref{integrand__non_qt} reduces to
\be
\label{cq3_em}
&&
-
\delta
\left(
r^2
\sqrt{\frac{h}{f}}
K^{[tr]}_{(\xi)}
\right)
-
r^2
\sqrt{\frac{h}{f}}
J^{[t}\xi^{r]}
\nonumber
\\
&=&
r^2
\sqrt{\frac{h}{f}}
\Big\{
-
\Big[
\frac{1}{8\pi G r}
+
\frac{A_0(r)A_0'(r)}{2h(r)}
+
2
\frac{g_4\left(X_0(r)\right)+2 f(r)\psi'(r)^2 g_{4,X}\left(X_0(r)\right) -f(r)^2 \psi'(r)^4 g_{4,XX}\left(X_0(r)\right)}
       {r}
\Big]
\delta f(r)
\nonumber\\
&&
+
\frac{f(r) A_0(r)A_0'(r)}{
2h(r)^2}
\delta h(r)
-
\frac{f(r)A_0(r)}{h(r)}
\delta A_0'(r)
+
{\cal J}^r
\delta\psi(r)
\nonumber
\\
&&+
\frac{4f(r)^2\psi'(r)}{r}
\left[
-g_{4,X}\left(X_0(r)\right)
+
f(r)\psi'(r)^2
g_{4,XX}\left(X_0(r)\right)
\right]
\delta\psi' (r)
\Big\},
\ee
where $X_0(r):=-\left(f(r)/2\right)\psi'(r)^2$ is the background value of the kinetic term $X$ and 
the nontrivial radial component of the Noether current associated with the shift symmetry is given by 
\be
{\cal J}^r
&=&
-
f(r) \psi'(r) 
g_{2,X}\left(X_0(r)\right)
+
\frac{2f(r)}{r^2h(r)}
\left[
\left(-1+f(r)\right)h(r)
+
rf(r) h'(r)
\right]
g_{4,X}\left(X_0(r)\right)
\nonumber\\
 &&
-
\frac{2f(r)^3\left(h(r)+rh(r)'\right)}
       {r^2 h(r)}
g_{4,XX}\left(X_0(r)\right).
\ee
We note that the model \eqref{irrational_f}
corresponds to a particular case of the general model \eqref{irrational_f2}.
Requiring that the background solution satisfies $f(r)=h(r)$, the equation of motion for $A_0(r)$ can be analytically integrated as
\be
A_0(r)=Q\left(\frac{1}{r}-\frac{1}{r_g}\right),
\ee
where the integration constant $Q$ represents the electric charge and we choose the integration constant to satisfy the gauge condition $A_0(r_g)=0$.
We assume the existence of the charged BHs that can be expanded near the event horizon $r=r_g$ as
\be
f(r)
=
h(r)
&=&
h_1 (r_g,Q) (r-r_g)
+
h_2(r_g,Q)(r-r_g)^2
+
{\cal O} 
\left[
(r-r_g)^3
\right],
\nonumber
\\
\psi(r)
&=&
\psi_{1/2}(r_g,Q)
\sqrt{r-r_g}
+
\psi_{3/2}(r_g,Q)
(r-r_g)^{\frac{3}{2}}
+
{\cal O} 
\left[
(r-r_g)^{\frac{5}{2}}
\right],
\ee
where the coefficients $h_i(r_g,Q)$ ($i=1,2,\cdots$) and $\psi_{j}(r_g,Q)$ ($j=1/2,3/2,\cdots$) are in general the functions of $r_g$ and $Q$, so that $X$ takes a nonzero constant value at the horizon
\be
X_0(r)
=
X_{0,0}
+ {\cal O} [r-r_g]
:=
-\frac{1}{8}h_1 \psi_{1/2}^2+ {\cal O} [r-r_g].
\ee
We require that the background solution satisfies ${\cal J}^r=0$,  so that the norm of the Noether current ${\cal J}^\mu {\cal J}_\mu=({\cal J}^r)^2/h(r)$ remains finite in the horizon limit $r\to r_g$, and then obtain at the leading order
\be
2r_g^2 
g_{2,X}(X_{0,0})
+
4\left(1-r_gh_1\right)
g_{4,X}(X_{0,0})
+
r_g h_1^2\psi_{1/2}^2
g_{4,XX}(X_{0,0})
=0.
\ee
The variation of the Hamiltonian on the horizon $r=r_g$ yields
\be
\label{var_horizon_general_f}
\delta{\cal H}_H
=
T_{\textsf{H}({\rm H})} \delta S_{\rm H}
&=&
\frac{r_g h_1}{2G}
\left[
1
+ 
16
\pi G g_4(X_{0,0})
+
4
\pi G h_1 \psi_{\frac{1}{2}}^2 g_{4,X}(X_{0,0})
\right]
\delta r_g,
\ee
where the Hawking temperature \eqref{hawking_temperature} is given by 
$T_{\textsf{H}({\rm H})} =h_1/(4\pi)$.
and hence 
\be
\label{proportional_f}
\delta S_{\rm H}
=
\frac{2\pi r_g}{G}
\left[
1
+ 
16
\pi G g_4(X_{0,0})
+
4
\pi G h_1 \psi_{\frac{1}{2}}^2 g_{4,X}(X_{0,0})
\right]
\delta r_g.
\ee
Since the proportionality coefficient in Eq.~\eqref{proportional_f} can depend on $Q$ in general, the differential~\eqref{proportional_f} may not be integrable ~\cite{Feng:2015wvb}.
However, since
\be
\frac{\partial}{\partial Q}
\left[
\frac{\delta S_{\rm H}}{\delta r_g}
\right]
=
-32\pi^2r_g
\left[
g_{4,X}(X_{0,0})
+
2X_{0,0}
g_{4,XX}(X_{0,0})
\right]
\frac{\partial X_{0,0}}{\partial Q},
\ee
there are two cases where the differential of the entropy is integrable.
The first case to satisfy the integrability condition $\frac{\partial}{\partial Q}\left[\delta S_{\rm H}/\delta r_g\right]=0$ is that
\be
\label{cond_integrability1}
g_{4,X}(X_{0,0})+2X_{0,0}g_{4,XX}(X_{0,0})=0.
\ee
The second case to satisfy $\frac{\partial}{\partial Q}\left[\delta S_{\rm H}/\delta r_g\right]=0$ is given by 
\be
\label{cond_integrability2}
\frac{\partial X_{0,0}}{\partial Q}=0,
\ee
namely, the kinetic term evaluated on the horizon $r=r_g$ does not depend on $Q$.
The condition \eqref{cond_integrability2} is essentially an extension of the result found for a particular choice of the $g_4(X)$ function, namely $g_4(X)=c' X$ with $c' $ being constant, discussed in Ref.~\cite{Feng:2015wvb}.
We note that the model discussed in Subsection \ref{sec6c} with $g_4(X)=\alpha\sqrt{-X}$ 
satisfies both the conditions \eqref{cond_integrability1} and \eqref{cond_integrability2},
since from Eq.~\eqref{bcl_solution_f2} we find that $X=-\frac{\alpha^2}{r^4\eta}$ does not depend on $Q$.

\section{Summary and conclusions}
\label{sec7}

We have investigated thermodynamics of static and spherically symmetric BHs in the Horndeski theories. 
Although the Wald entropy formula has been useful for computing the BH entropy in the covariant gravitational theories which contain the dependence only on the Riemann tensors as the higher-derivative terms, this may not be directly applicable to the Horndeski theories because of the presence of the derivative interactions and the nonminimal derivative couplings of the scalar field to the spacetime curvature tensors.
The terms which contain the spacetime curvature tensors may be eliminated with use of the properties of the Riemann tensor, and the apparent dependence of the action on the spacetime curvatures may be modified before and after a partial integration.
Thus, following the original formulation by Iyer and Wald, we have employed the Noether charge potential associated with the differmorphism invariance.
The variation of the Noether charge potential on the boundaries is related to the variation of the Hamiltonian.
The variations of the Hamiltonian on the BH event horizon and at the spatial infinity, respectively, give rise to the differentials of the entropy of the BH and the total mass of the system, and the conservation of the total Hamiltonian leads to the first law of the BH thermodynamics.
Our formulation could be applied to the whole of the Horndeski theories including the EsGB theories and the shift-symmetric theories which provide the stealth Schwarzschild BH solutions with the linearly time-dependent scalar field.
In the case of the EsGB theories, our formulation has recovered the standard Wald entropy formula, although the description of the EsGB theory in the context of the Horndeski theories appears to be different from the original action by the difference in the total derivative terms.

We have divided our analysis into the two parts.
The first part is about the static and spherically symmetric BH solutions with the static scalar field in the Horndeski theories which may not be shift symmetric.
The second part is about those with the linearly time-dependent scalar field in the shift-symmetric Horndeski theories.
In the latter case, in order to satisfy the radial-temporal component of the gravitational equations, the radial component of the Noether current associated with the shift symmetry has to vanish.
Taking this into consideration, we showed that the variation of the Noether charge potential associated with the diffeomorphism invariance does not depend on time, even if the scalar field has a linear time dependence.
This reflects the fact that in such static and spherically symmetric BH solutions there was no radial heat flux onto the BH horizon.

The results in the former part are summarized in Table~\ref{Table:summary1}.
\begin{table}[htbp]
\begin{center}
\scalebox{0.8}[0.9]{
  \begin{tabular}{|c|c|c||c|c||c|c|c|}
\hline 
\multicolumn{3}{|c||}{Theory}&BH &Scalar field& Temperature $T_{\textsf{H}}$& Mass $M$& Entropy $ S$
 \\
\hline
\hline
 \lower2ex\hbox{I}& 
\multicolumn{2}{c||}{GR without $\Lambda$}  
&
Schwarzschild
&
trivial
&
$T_0:=\frac{1}{4\pi r_g}$
& 
$M_0:=\frac{r_g}{2G}$
&
$S_0:= \frac{\pi r_g^2}{G}$
\\[-.5ex]
\cline{2-8}
\\[-1em]
&\multicolumn{2}{c||}{GR with $\Lambda$}
&
Schwarzschild-(A)dS
&
trivial
&
$\left(1-\Lambda r_g^2\right)
T_0$
& 
$
\left(1-\frac{\Lambda}{3} r_g^2\right)
M_0
$
&
$S_0$
\\
 \hline \hline
 \\[-1em]
II &\multicolumn{2}{c||}{conventional Scalar-Tensor  theories }
&
Schwarzschild-(A)dS
&
$\phi=0$
&
$\left(1-V(0)r_g^2\right)
T_0
$
& 
$
\left(1-\frac{V(0)}{3} r_g^2\right)
M_0
$
&
$S_0$
\\
\hline
 & \lower4ex\hbox{
non-shift-symmetric
 EsGB} &\lower2ex\hbox{$k^{(1)}(\phi_0)=0$}
 &
asymptotically-flat
& 
hairy
&
$\frac{1}{4\pi}\sqrt{f_1 h_1}$\,(Eqs.~\eqref{h_exp}-\eqref{phi_exp})
&
{\rm ADM mass}
& 
$\left(1+\frac{64
\pi Gk[\psi_{H}(r_g)]}{r_g^2}\right)
S_0
$
\\[-3ex]
\cline{4-8}
III&&&
Schwarzschild
&
$\phi=\phi_0$
&
$T_0$
&
$M_0$
&
$S_0$
\\
\cline{3-8}
& 
 &
$k^{(1)}(\phi)\neq 0$
&
asymptotically-flat
& 
hairy
&
$\frac{1}{4\pi}\sqrt{f_1 h_1}$\,(Eqs.~\eqref{h_exp}-\eqref{phi_exp})
&
{\rm ADM mass}
& 
$\left(1+\frac{64
\pi Gk[\psi_{H}(r_g)]}{r_g^2}\right)
S_0
$
\\
\cline{2-8}
& 
\multicolumn{2}{c||}{shift-symmetric
EsGB}
&
asymptotically-flat
& 
hairy
&
$\frac{1}{4\pi}\sqrt{f_1 h_1}$\,(Eqs.~\eqref{h_exp}-\eqref{phi_exp})
&
{\rm ADM mass}
& 
$S_0$
\\
 \hline
IV&\multicolumn{2}{c||}{Horndeski with $G_4=\alpha\sqrt{-X}$}&
asymptotically (A)dS
&
hairy
&
$
\left(
1-\Lambda r_g^2
+
\frac{8\pi G \alpha^2}{\eta r_g^2}
\right)
T_0$
& 
$
\left(
1-\frac{\Lambda }{3}r_g^2-\frac{8\pi G \alpha^2}{\eta r_g^2}
\right)
M_0
$
&
$S_0$
\\
\hline
 \end{tabular}
 }
\caption{Thermodynamical properties of BHs with/without static scalar field are summarized.
In the case of non-shift-symmetric EsGB BH (Theory III), if $k^{(1)}(\phi_0)=0$ where $\phi_0$ is some constant, there exists a trivial Schwarzschild BH, while when $k^{(1)}(\phi)\neq 0$ for any real $\phi$, non-trivial solution is unique.
Note that we can make $\phi_0=0$ after a suitable shift of $\phi$.
The former class of non-shift-symmetric Theory III includes scalarized BH solutions for the $Z_2$-symmetric coupling models \cite{Silva:2017uqg,Doneva:2017bvd,Antoniou:2017acq,Blazquez-Salcedo:2018jnn,Minamitsuji:2018xde,Silva:2018qhn,Cunha:2019dwb,Konoplya:2019fpy,Doneva:2021dqn,East:2021bqk,Julie:2022huo,Doneva:2020nbb,Doneva:2022yqu,Dima:2020yac,Herdeiro:2020wei,Lai:2023gwe}, while the latter of the non-shift-symmetric Theory III includes hairy BH solutions in the exponential coupling models~\cite{Kanti:1995vq,Kanti:1997br,Torii:1996yi}.
The shift-symmetric class of Theory III corresponds to the linear coupling model~\cite{Sotiriou:2013qea,Sotiriou:2014pfa}.
Hawking temperature $T_{\textsf{H}}$, mass $M$ and entropy $S$ of non-trivial EsGB BH are given by numerical solutions.
Theory I (GR with/without $\Lambda$) is also listed as a reference.
}
\label{Table:summary1}
\end{center}
\end{table}
Besides GR and the conventional ST theory with the trivial scalar field, we evaluated the BH entropy and the total mass of the system for the static and spherically symmetric BHs with nontrivial profile of the scalar field in the shift-symmetric EsGB theory and in the shift-symmetric theory where the function $G_4(X)$ contains the term proportional to $\sqrt{-X}$.
In both cases, we showed that the BH entropy was given by the area law despite the existence of the nontrivial profile of the scalar field.

The results in the latter part are summarized in Table~\ref{Table:summary2}.
We have studied the BH entropy and the mass in the stealth Schwarzschild solution and the Schwarzschild-(A)dS solution with the linearly time dependent scalar field.
In both cases, we have found that the BH entropy does not obey the area law and the total mass of the system does not coincide with the BH mass from the metric.
\begin{table}[htbp]
\begin{center}
\scalebox{0.8}[0.9]{
  \begin{tabular}{|c|c||c|c||c|c|c|}
\hline 
\multicolumn{2}{|c||}{Theory}&BH &Scalar field&Temperature $T_{\textsf{H}}$& Mass $M$& Entropy $ S$
 \\
\hline
&
\lower1ex\hbox{$G_2(X),\, G_4(X)$ with}
&
&
&
&
&
\\
V &
\raise.5ex\hbox{$G_2
\left(\frac{q^2}{2}\right)
=
G_{2X}
\left(\frac{q^2}{2}\right)=0$}
 &
\raise2ex\hbox{Schwarzschild}
& 
\raise2ex\hbox{hairy}
& 
\raise2ex\hbox{
$T_0$
}
& 
\raise2ex\hbox{$
\left(1
-
\frac{G_{4X}q^2}{G_4}\right)
M_0$}
&
\raise2ex\hbox{$
\left(1
-
\frac{G_{4X}q^2}{G_4}\right)
S_0
$
}
\\
\cline{2-7}
&
&
& 
\lower1ex\hbox{hairy}
& 
\lower1ex\hbox{$T_0$}
& 
\lower1ex\hbox{$
\left(1
-
8
\pi G
q^2\beta \right)M_0
$}
&
\lower1ex\hbox{$
\left(1
-
8\pi Gq^2\beta \right)
S_0
$}
\\
\cline{4-7}
&
\raise2ex\hbox{$G_4(X)=
\frac{1}{16\pi G}
+\beta X$}
&
\raise2ex\hbox{Schwarzschild}
&\raise1ex\hbox{trivial}
& 
\raise1ex\hbox{$T_0$}
& 
\raise1ex\hbox{$
M_0
$}
&
\raise1ex\hbox{
$S_0$
}
\\
\hline
&
&
&
\lower.5ex\hbox{hairy}
&
&
&
\\
VI &
\raise2ex\hbox{$G_2(X)=\eta X-\frac{1}{8\pi G \Lambda}$},
&
Schwarzschild-(A)dS
&
$q=\sqrt{\frac{\eta+2\beta \Lambda}
       {16\pi G \beta \eta}}$
& 
\raise2ex\hbox{$\left(1-\bar\Lambda r_g^2\right)T_0$}
& 
\raise2ex\hbox{$\left(1
-
8\pi Gq^2\beta \right)\left(1-
\frac{\bar\Lambda}{3}
r_g^2\right)M_0$}
&
\raise2ex\hbox{$\left(1
-
8\pi Gq^2\beta \right)
S_0$}
\\
\cline{4-7}
&
\raise2ex\hbox{$G_4(X)=
\frac{1}{16\pi G}
 +\beta X$}
 &&\raise.5ex\hbox{trivial}& 
\raise.5ex\hbox{$\left(1-\Lambda r_g^2\right)T_0$}
& 
\raise.5ex\hbox{$
\left(1-
\frac{\Lambda}{3}
r_g^2\right)M_0$}
&
\raise.5ex\hbox{
$S_0$
}
\\
\hline
\lower2ex\hbox{VII} &
\lower1ex\hbox{$
G_2(X)\,,
G_3(X)
$}
&
&
&
&
&
\\
&
with $c_{\rm gw}^2=1$
&
\raise2ex\hbox{Schwarzschild}
&
\raise2ex\hbox{hairy}
& 
\raise2ex\hbox{$T_0$}
& 
\raise2ex\hbox{$M_0$}
&
\raise2ex\hbox{
$S_0$}
\\
\hline
 \end{tabular}
 }
\caption{Thermodynamical properties of BHs with linearly time-dependent scalar field in shift-symmetric Horndeski theories are summarized. The scalar field has linear-time dependence as $\phi=qt+\psi(r)$.
In Theory VI, since $\bar \Lambda:= -\frac{\eta}{2\beta}$, we find $1-8\pi Gq^2\beta=\frac{1}{2}\left(1+\frac{\Lambda}{\bar \Lambda} \right)$. 
$T_0$, $M_0$ and $S_0$ are defined in Table~\ref{Table:summary1}.}
\label{Table:summary2}
\end{center}
\end{table}

In Theory V and Theory VI in Table~\ref{Table:summary2}, there exists a trivial Schwarzschild solution without scalar field. 
Then we have discussed the thermodynamic stability of the stealth Schwarzschild BHs. We have shown that its stability depends on the sign of the nonminimal derivative coupling to the spacetime curvature.
In the case of the Schwarzschild-dS BH, we have shown the Horndeski Schwarzschild-dS BHs are always thermodynamically unstable  and transit to the GR Schwarzschild-dS BH.
In the case of the Schwarzschild-AdS BHs, we have found that the thermodynamical phase diagram becomes more complicated than the previous case, because of the existence of the Hawking-Page phase transition, and crucially depends on the ratio of the (effective) cosmological constants, i.e., the ratio of  the Horndeski AdS radius ${\bar \ell}$ and the GR AdS radius ${\ell}$, where we always have ${\bar\ell}<\ell$.
We have shown that in the case that ${\bar \ell}$ is not much less than ${\ell}$, the Horndeski Schwarzschild-AdS BH is always thermodynamically unstable and decays into either the GR Schwarzschild-AdS BH or the AdS spacetime filled with thermal radiation.
On the other hand, in the case that the ratio of ${\bar \ell}$ to ${\ell}$ is smaller than a critical value, there is a certain range of the BH mass where the Horndeski Schwarzschild-AdS BH is thermodynamically more stable than the GR Schwarzschild-AdS BH, while for the BH mass larger than that in this range the Horndeski Schwarzschild-AdS BH decays into either the GR Schwarzschild-AdS BH or the AdS spacetime with thermal radiation.

While BH solutions discussed so far contain only one independent charge, i.e., the mass or  equivalently the horizon radius, in Section VI we have briefly discussed thermodynamics in the BHs with two independent charges in the Horndeski theories.
More concretely, we have focused on the Horndeski theories minimally coupled to the $U(1)$-invariant vector field, where BH solutions contain the two independent charges, the mass and the electric charge.
By extending the general formulation presented in Section II, we have derived the Noether change potential associated with the diffeomorphism invariance, including the contribution of the $U(1)$-invariant vector field with the nonlinear kinetic term. 
As a concrete example of charged BH solutions in the Horndeski theories, we have considered the extension of the irrational coupling model discussed in Sec.~\ref{sec3d} minimally coupled to the Maxwell field, and showed that in spite of the presence of the two independent charges the differential of the entropy is integrable and the ordinary area law is recovered.
Finally, in the general reflection- and shift-symmetric class of the Horndeski theories with the minimally coupled $U(1)$-invariant vector field, we have clarified the general conditions under which the differential of the BH entropy is integrable in the presence of the two independent charges.
We have shown that in the case that the kinetic term of the scalar field evaluated on the horizon does not depend on the electric charge the differential of the BH entropy is integrable.

There would be various extensions of our present work, which include the cases of the stationary and axisymmetric BHs in the Horndeski theories and the nontrivial BHs in the healthy ST theories beyond the Horndeski theories \cite{Langlois:2015cwa,BenAchour:2016fzp,Takahashi:2021ttd}. We hope to come back to these cases in our future work.

\section*{ACKNOWLEDGMENTS}
M.M.~was supported by the Portuguese national fund through the Funda\c{c}\~{a}o para a Ci\^encia e a Tecnologia in the scope of the framework of the Decree-Law 57/2016 of August 29, changed by Law 57/2017 of July 19, and the Centro de Astrof\'{\i}sica e Gravita\c c\~ao through the Project~No.~UIDB/00099/2020. K.M. would like to acknowledges the Yukawa Institute for Theoretical Physics at Kyoto University, where the present work was begun during the Visitors Program of FY2021. He would also thank CENTRA/Instituto Superior T\'ecnico, where some of this work was performed during his stay. This work was supported in part by JSPS KAKENHI Grant Numbers  JP17H06359, JP19K03857.
\appendix

\bibliographystyle{apsrev4-1}
\bibliography{refs}
\end{document}